%% file: main.tex
\documentclass[draftclsnofoot,journal,onecolumn, 12pt]{IEEEtran}
\usepackage{subfiles}
\usepackage{amsmath,graphicx,dsfont}
\usepackage{mathtools}
\usepackage{algorithm}
\usepackage{algpseudocode}
\input{preamble_arxiv.tex}

\graphicspath{{./Figures/}}

\newcommand{\ARL}{\text{ARL}}

\newcommand{\tS}{\widetilde{S}_N}

\newcommand{\wmu}[1]{\widehat{g}_N^{#1}}
\usepackage{xr}
\begin{document} 
	\subfile{BG_CuSum_v29_single_column_arxiv.tex}

\subfile{supplementary_materials_arxiv.tex}
\end{document}

%% file: preamble_arxiv.tex
\usepackage[T1]{fontenc}
\usepackage{amsmath,amssymb,amsfonts,mathrsfs,bm}% Typical maths resource packages
\usepackage{amsthm}
\usepackage{cite}
\usepackage{array}
\usepackage[shortlabels]{enumitem}
\usepackage{graphicx}
\usepackage{url}
\usepackage{color}
\usepackage{multirow}
\usepackage[table]{xcolor}
\usepackage[normalem]{ulem}
\usepackage{array}
\newcolumntype{L}[1]{>{\raggedright\let\newline\\\arraybackslash\hspace{0pt}}m{#1}}
\newcolumntype{C}[1]{>{\centering\let\newline\\\arraybackslash\hspace{0pt}}m{#1}}
\newcolumntype{R}[1]{>{\raggedleft\let\newline\\\arraybackslash\hspace{0pt}}m{#1}}
\usepackage{xparse}

\usepackage[hidelinks]{hyperref} %load as the last package

\ifx\useTheoremCounter\undefined
\newtheorem{Theorem}{Theorem}
\newtheorem{Corollary}{Corollary}
\newtheorem{Proposition}{Proposition}
\newtheorem{Lemma}{Lemma}
\else
\newtheorem{Theorem}{Theorem}

\newtheorem{Proposition}[Theorem]{Proposition}

\fi

\newtheorem{Definition}{Definition}

\newtheorem{Assumption}{Assumption}

\newtheorem{Theorem_A}{Theorem}[section]
\newtheorem{Proposition_A}{Proposition}[section]
\newtheorem{Lemma_A}{Lemma}[section]
\newtheorem{Corollary_A}{Corollary}[section]

\theoremstyle{remark}

\newcommand{\Real}{\mathbb{R}}
\newcommand{\Nat}{\mathbb{N}}

\newcommand{\calN}{\mathcal{N}}

\newcommand{\bbQ}{\mathbb{Q}}

\DeclareSymbolFont{bsfletters}{OT1}{cmss}{bx}{n}
\DeclareSymbolFont{ssfletters}{OT1}{cmss}{m}{n}
\DeclareMathSymbol{\bsfGamma}{0}{bsfletters}{'000}
\DeclareMathSymbol{\ssfGamma}{0}{ssfletters}{'000}
\DeclareMathSymbol{\bsfDelta}{0}{bsfletters}{'001}
\DeclareMathSymbol{\ssfDelta}{0}{ssfletters}{'001}
\DeclareMathSymbol{\bsfTheta}{0}{bsfletters}{'002}
\DeclareMathSymbol{\ssfTheta}{0}{ssfletters}{'002}
\DeclareMathSymbol{\bsfLambda}{0}{bsfletters}{'003}
\DeclareMathSymbol{\ssfLambda}{0}{ssfletters}{'003}
\DeclareMathSymbol{\bsfXi}{0}{bsfletters}{'004}
\DeclareMathSymbol{\ssfXi}{0}{ssfletters}{'004}
\DeclareMathSymbol{\bsfPi}{0}{bsfletters}{'005}
\DeclareMathSymbol{\ssfPi}{0}{ssfletters}{'005}
\DeclareMathSymbol{\bsfSigma}{0}{bsfletters}{'006}
\DeclareMathSymbol{\ssfSigma}{0}{ssfletters}{'006}
\DeclareMathSymbol{\bsfUpsilon}{0}{bsfletters}{'007}
\DeclareMathSymbol{\ssfUpsilon}{0}{ssfletters}{'007}
\DeclareMathSymbol{\bsfPhi}{0}{bsfletters}{'010}
\DeclareMathSymbol{\ssfPhi}{0}{ssfletters}{'010}
\DeclareMathSymbol{\bsfPsi}{0}{bsfletters}{'011}
\DeclareMathSymbol{\ssfPsi}{0}{ssfletters}{'011}
\DeclareMathSymbol{\bsfOmega}{0}{bsfletters}{'012}
\DeclareMathSymbol{\ssfOmega}{0}{ssfletters}{'012}

\DeclareMathOperator*{\argmax}{arg\,max}

\DeclareMathOperator*{\esssup}{ess\,sup}

\newcommand{\qednew}{\nobreak \ifvmode \relax \else
      \ifdim\lastskip<1.5em \hskip-\lastskip
      \hskip1.5em plus0em minus0.5em \fi \nobreak
      \vrule height0.75em width0.5em depth0.25em\fi}

\newcommand{\ud}{\mathrm{d}}

\newcommand{\ofrac}[1]{{\frac{1}{#1}}}

\newcommand{\ceil}[1]{\left\lceil{#1}\right\rceil}
\newcommand{\floor}[1]{\left\lfloor{#1}\right\rfloor}

\newcommand{\KLD}[2]{{D_{\text{KL}}({#1}\ \|\ {#2})}}

\newcommand{\cond}[2]{\left. {#1}\, \middle| \, {#2} \right.}

\DeclareDocumentCommand \P { g d() g } {%
	\IfNoValueTF {#3}
	{%
		\IfNoValueTF {#1}
		{%
			\IfNoValueTF {#2}
			{%
				\mathbb{P}%
			}%
			{%
				\mathbb{P}\left({#2}\right)%
			}%
		}%
		{%
			\IfNoValueTF {#2}
			{%
				\mathbb{P}_{#1}%
			}%
			{%
				\mathbb{P}_{#1}\left({#2}\right)%
			}%		
		}%
	}%
	{%
		\IfNoValueTF {#1}
		{%
			\mathbb{P}\left(\cond{#2}{#3}\right)%
		}%
		{%
			\mathbb{P}_{#1}\left(\cond{#2}{#3}\right)%
		}%	
	}%
}

\DeclareDocumentCommand \E { g o g } {%
	\IfNoValueTF {#3}
	{%
		\IfNoValueTF {#1}
		{%
			\IfNoValueTF {#2}
			{%
				\mathbb{E}%
			}%
			{%
				\mathbb{E}\left[{#2}\right]%
			}%
		}%
		{%
			\IfNoValueTF {#2}
			{%
				\mathbb{E}_{#1}%
			}%
			{%
				\mathbb{E}_{#1}\left[{#2}\right]%
			}%		
		}%
	}%
	{%
		\IfNoValueTF {#1}
		{%
			\mathbb{E}\left[\cond{#2}{#3}\right]%
		}%
		{%
			\mathbb{E}_{#1}\left[\cond{#2}{#3}\right]%
		}%	
	}%
}

\definecolor{gray90}{gray}{0.9}

\newcommand{\hide}[1]{}
\newcommand{\figref}[1]{\figurename~\ref{#1}}
\renewcommand{\figurename}{Fig.}
\graphicspath{{./Figures/}}

%% file: BG_CuSum_v29_single_column_arxiv.tex
	\maketitle
	\begin{abstract}
		The problem of quickest detection of a change in distribution is considered under the assumption that the pre-change distribution is known, and the post-change distribution is only known to belong to a family of distributions distinguishable from a discretized version of the pre-change distribution. A sequential change detection procedure is proposed that partitions the sample space into a finite number of bins, and monitors the number of samples falling into each of these bins to detect the change. A test statistic that approximates the generalized likelihood ratio test is developed. It is shown that the proposed test statistic can be efficiently computed using a recursive update scheme, and a procedure for choosing the number of bins in the scheme is provided. Various asymptotic properties of the test statistic are derived to offer insights into its performance trade-off between average detection delay and average run length to false alarm. Testing on synthetic and real data demonstrates that our approach is comparable or better in performance to existing non-parametric change detection methods.
	\end{abstract}
	\section{Introduction}
	\label{sec:intro}
	Quickest change detection (QCD) is a fundamental problem in statistics. Given a sequence of observations that have a certain distribution up to an unknown change point $\nu$, and have a different distribution after that, the goal is to detect this change in distribution as quickly as possible subject to false alarm constraints. The QCD problem arises in many practical situations, including applications in manufacturing such as quality control, where any deviation in the quality of products must be quickly detected. With the increase in the amount and types of data modern-day sensors are able to observe, sequential change detection methods have also found applications in the areas of power system line outage detection\cite{chen2016quickest,rovatsos2017statistical}, bioinformatics\cite{muggeo10}, network surveillance\cite{akoglu10,sequeira02,Mar10,androulidakis06,androulidakis08,wang04}, fraud detection\cite{bolton02}, structural health monitoring\cite{sohn00}, spam detection\cite{xie12}, spectrum reuse \cite{Lai2008,kundargi09,ibrahim14,sahasranand15,Zhang2016,Hanafi2016,cheng16}, video segmentation\cite{li04}, and resource allocation and scheduling \cite{Tang2014,Ren2017}. In many of these applications, the detection algorithm has to operate in real time with reasonable computation complexity.
	
	% Setting and Contributions
	In this paper, we consider the QCD problem with a known pre-change distribution and unknown post-change distribution. The only information we have about the post-change distribution is that it belongs to a set of distributions that is distinguishable from the pre-change distribution when the sample space is discretized into $N$ bins. Assuming that the pre-change distribution is known is reasonable, because in most practical applications, a large amount of data generated by the pre-change distribution is available to the observer who may use this data to obtain an accurate approximation of the pre-change distribution \cite{escobar95,li99}. However, estimating or even modelling the post-change distribution is often impractical as we may not know \emph{a priori} what kind of change will happen. We seek to design a low-complexity detection algorithm that allows us to quickly detect the change, under false alarm constraints, and with minimal knowledge of the post-change distribution. To solve this problem, we propose a new test statistic based on binning and the generalized likelihood ratio test (GLRT), which approximates the Cumulative Sum  (CuSum) statistic of Page \cite{page54}. We propose a method to choose the appropriate number of bins, and show that our proposed test statistic can be updated sequentially in a manner similar to the CuSum statistic. We also provided an analysis of the performance of the proposed test statistic.
	
	\subsection{Related Works}
	
	For the case where the pre- and post-change distributions are known, Page \cite{page54} developed the CuSum Control Chart for QCD, which was later shown to be asymptotically optimal under a certain criterion by Lorden\cite{lorden71}, and exactly optimal under Lorden's criterion by Moutakides\cite{moustakides86}. We refer the reader to \cite{poor2009quickest,tartakovsky2014sequential,veeravalli2013quickest} and the references therein for an overview of the QCD problem. There are also many existing works that consider the QCD problem where the post-change distribution is unknown to a certain degree. In \cite{siegmund95}, the authors considered the case where the post-change distribution belongs to a one-parameter exponential family with the pre-change distribution being known. The case where both the pre- and post-change distributions belong to a one-parameter exponential family was considered by \cite{lai98}. In \cite{banerjee15}, the authors developed a data-efficient scheme that allows for optional sampling of the observations in the case when either the post-change family of distributions is finite, or both the pre- and post-change distributions belong to a one parameter exponential family. In \cite{pawlak13}, the authors assumed that the observations are time-dependent as in ARMA, general linear processes and $\alpha$-mixing processes. In \cite{sakurai12}, the authors proposed using a infinite hidden Markov model for tracking parametric changes in the signal model. Classical approaches to the QCD problem without strong distributional assumptions can be found in \cite{gordon94,hawkins10,rafajlowicz10,rafajlowicz09}. Although there are no distributional assumptions, the type of change assumed in \cite{gordon94,rafajlowicz09,hawkins10,rafajlowicz10,lang15} is a shift in the mean and in \cite{ross11}, a shift in the scale of the observations. In \cite{li15}, the authors provided a kernel-based detection scheme for a change-point detection problem where the post-change distribution is completely unknown. However, the kernel-based detection scheme requires choosing a proper kernel bandwidth. This is usually done by tuning the bandwidth parameter on a reference dataset representative of the pre-change and possible post-change conditions. In \cite{darkhovskii1988nonparametric}, the authors proposed a non-parametric algorithm for detecting a change in mean. They assume that the mean of the pre- and post-change distributions are not equal but do not assume any knowledge of the distribution function. 
	
	In this paper, we consider the case where less information is available about the post-change distribution, compared to \cite{siegmund95,lai98,banerjee15}. The only assumption we make is that the post-change distribution belongs to a set of distributions that are distinguishable from the pre-change distribution when the sample space is discretized into a known number of bins. Our method is also more general than \cite{gordon94,hawkins10,ross11,darkhovskii1988nonparametric, rafajlowicz09,rafajlowicz10,lang15} 
	as it is not restricted to a shift in mean, location or scale. 
	Related work is discussed in \cite{nitinawarat15}, in which an asymptotically optimal universal scheme is developed to isolate an outlier data stream that experiences a change, from a pool of at least two typical streams. In our work, we only have a single stream of observations and therefore our scheme is unable to make use of comparisons across streams to detect for changes. Other related works include \cite{li15,ross12,hawkins10}, which do not require prior information about the pre-change distribution. These works first partition the observations using a hypothesized change point into two sample sets and then test if they are generated by the same distribution using a two-sample test. This approach results in weak detection performances when the change occurs early in the observation sequence. In our work, we assume that prior information about the pre-change distribution is available, which allows us to detect the change even when the change occurs early in the observations. 
	
	\subsection{Our Contributions}
	In this paper, we consider the QCD problem where the pre-change distribution is known, and the post-change distribution belongs to a family of distributions distinguishable from the pre-change distribution when the sample space is discretized into bins. Our goal is to develop an algorithm that is effective and has low computational complexity. Our main contributions are as follows:
	\begin{enumerate}
		\item We propose a Binned Generalized CuSum (BG-CuSum) test and derive a recursive update formula for our test statistic, which allows for an efficient implementation of the test. 
		%\item We present properties of the family of post-change distributions distinguishable from the pre-change distribution, and indicate how the number of bins can be chosen.
		\item We derive a lower bound for the average run length (ARL) to false alarm of our BG-CuSum test, and show asymptotic properties of the BG-CuSum statistic in order to provide insights into the trade-off between the average detection delay (ADD) and ARL. 
		\item We provide simulations and experimental results, which indicate that our proposed BG-CuSum test outperforms various other non-parametric change detection methods and its performance approaches a test with known asymptotic properties as the ARL becomes large.
	\end{enumerate}
	A preliminary version of this work was presented in \cite{lau17}. To the best of our knowledge, the only other recursive test known in the literature for the case where the post-change distribution is not completely specified and does not belong to a finite set of distributions is discussed in \cite{lorden2008sequential}. The paper \cite{lorden2008sequential} derived a scheme that uses 3 registers for storing past information to detect a change in the parameter of an exponential family. Our work is applicable to more general changes in distribution as we do not assume that the post-change distribution belong to an exponential family. Due to technical difficulties, we are not able to obtain the asymptotic ADD for our BG-CuSum test. Instead, we derive the asymptotic ADD for a related non-recursive test, which is asymptotically optimal when the pre- and post-change distributions are discrete. Simulations indicate that the BG-CuSum test has similar asymptotic behavior as this latter test, which is however computationally expensive. The asymptotic performance analysis of the BG-CuSum test remains an open problem.
	
	The rest of this paper is organized as follows. In Section \ref{sec:problem}, we present the QCD signal model and problem formulation. In Section \ref{sec:algorithm}, we present our GLRT based QCD procedure and propose an approximation that can be computed efficiently. %In Section \ref{sec:distinguishable_distributions}, we present some properties of the family of distributions distinguishable from the pre-change distribution. 
	In Section \ref{sec:asymptotics}, we analyse the asymptotic behavior of our test statistic, and provide a heuristic analysis of the trade-off between the ADD and the ARL to false alarm. In Section \ref{sec:results}, we present simulation and experimental results to illustrate the performance of our algorithm. Finally, in Section \ref{sec:conclude}, we present some concluding remarks.
	
	\emph{Notations:} We use $\Real$ and $\Nat$ to denote the set of real numbers and positive integers, respectively. The operator $\E^f$ denotes mathematical expectation with respect to (w.r.t.) the probability distribution with generalized probability density (pdf) $f$, and $X \sim f$ means that the random variable $X$ has distribution with pdf $f$. We use upper-case (e.g. $X$) to refer to a random variable, and lower case (e.g. $x$) to refer to a realization of the random variable $X$. 	%<*tag:r1c2>
	We let $\P{\nu}$ and $\E_\nu$ denote the measure and mathematical expectation respectively where the change point is at $\nu$ for $\nu\in\mathbb{N}$. In particular, $\P{1}$ and $\E_1$ denote the measure and mathematical expectation, respectively, when the change has occurred at time $t=1$, i.e., all the observations are distributed according to the post-change distribution.
	%</tag:r1c2>
	Similarly, we let and $\P{\infty}$ and $\E{\infty}$ denote the measure and mathematical expectation, respectively, when there is no change, i.e., all the observations are distributed according to the pre-change distribution. The Gaussian distribution with mean $\mu$ and variance $\sigma^2$ is denoted as $\calN(\mu,\sigma^2)$. The Dirac delta function at $\theta$ is denoted as $\delta_\theta$. $|A|$ denotes the cardinality of the set $A$.

	\section{Problem formulation}\label{sec:problem}
	
	We consider distributions defined on the sample space $\Real$. Let ${f}=p_0f_c+\sum_{h=1}^Hp_h\delta_{\theta_h}$ be the generalized pdf of the pre-change distribution, where $f_c$ is the pdf of the continuous part, $\{\theta_h \in \Real : h=1,\ldots,H\}$ are the locations of the point masses, and $0\leq p_h \leq 1$ for all $h\geq 0$ with $\sum_{h=0}^H p_h =1$. Similarly, let $g=q_0g_c+\sum_{h=1}^Hq_h\delta_{\theta_h}$ be the generalized pdf of the post-change distribution where $g\neq f$, and $0\leq q_h \leq 1$ for all $h\geq 0$ with $\sum_{h=0}^H q_h=1$. Let $X_1,X_2,\ldots$ be a sequence of real valued random variables satisfying the following:
	\begin{align}\label{eqn:signalmodel}
		\begin{cases}
			X_t \sim {f} \quad \text{i.i.d.\ for all $t< \nu$},\\
			X_t \sim g \quad \text{i.i.d.\ for all $t\geq \nu$},\\
		\end{cases}
	\end{align}
	where $\nu\geq 0$ is an unknown but deterministic change point. The quickest change detection problem is to detect the change in distribution, through observing $X_1=x_1,X_2=x_2,\ldots$, as quickly as possible while keeping the false alarm rate low. In this paper, we assume that the observer only knows the pre-change distribution ${f}$ but does not have full knowledge of the post-change distribution $g$. We assume that the post-change distribution is distinguishable from the pre-change distribution when sample space is discretized into a known number of bins, defined as follows.
	
	\begin{Definition}\label{def:bins}	
		Let $\Theta=\{\theta_1,...,\theta_H\}$. For any $N\geq 1$, let $I^N_1=(-\infty,z_1]\backslash \Theta$, $I^N_2=(z_1,z_2]\backslash \Theta$, $\ldots$, and $I^N_N=(z_{N-1},\infty)\backslash \Theta$ be $N$ sets such that for each $j\in\{1,...,N\}$, we have $\int_{I^N_j} f_c(x) \ud x=\frac{1}{N}.$ Also define the sets $I^N_{N+1}=\{\theta_1\}, \ldots, I^N_{N+H}=\{\theta_H\}$. We call each of the sets $I^N_j, j=1,\ldots,N+H$, a bin. 
	\end{Definition}
	
	\begin{Definition}\label{def:distinguishable}
		A distribution $g$, absolutely continuous w.r.t. $f$, is distinguishable from $f$ w.r.t.\ $N$ if there exists $j\in\{1,...,N+H\}$ such that $\int_{I^N_j} f(x) \ud x \neq \int_{I^N_j} g(x) \ud x$. The set $D(f,N)$ is the family of all generalized pdfs $g$ distinguishable from $f$ w.r.t.\ $N$.
	\end{Definition}
	
	Any distribution $g\ne f$ is distinguishable from $f$ for $N$ sufficiently large. To see why this is true, we define the functions $g_N$ and $f_N$ as
	\begin{align}
		g_N(x)\doteq\int_{I^N_j} g(y)\ \ud y,\quad
		f_N(x)\doteq \int_{I^N_j} f(y)\ \ud y, \label{notation_abuse}
	\end{align}
	where $j$ is the unique integer such that $x\in I^N_j$. If there exists $h\in\{0,...,H\}$ such that $p_h\neq q_h$ then $g\in D(f,1)$. Now suppose that $p_h=q_h$ for all $h\in\{0,...,H\}$. Let $F_c$ and $G_c$ be the cumulative distribution function of ${f_c}$ and $g_c$, respectively. Since both $F_c$ and $G_c$ are continuous and $F_c\neq G_c$, there exists an interval $J$ such that for any $x\in J$, $F_c(x)\ne G_c(x)$. Then for any $N > 1/ \int_J f_c(x) \ud x$, there exists some $x$ such that $g_N(x) \ne f_N(x)$ because otherwise $G_c(x)=F_c(x)$ for any $x$ that is a boundary point of a set in $\{I^N_j: j=1,\ldots,N\}$. 
	
	In the supplementary material \cite{supplementary}, we give an exact characterization of $N$ such that a distribution $g$ is distinguishable from $f$ w.r.t. $N$. We also provide an example to illustrate how $N$ can be determined if additional moment information is available. For the rest of this paper, we make the following assumption.
	
	\begin{Assumption}
		For a known positive integer $N$ and pre-change distribution $f$, the post-change distribution $g$ belongs to the set $D(f,N)$. 
	\end{Assumption}	
	
	In a typical sequential change detection procedure, at each time $t$, a test statistic $S(t)$ is computed based on the currently available observations $X_1=x_1,\ldots,X_t=x_t$, and the observer makes the decision that a change has occurred at a stopping time $\tau$, where $\tau$ is the first $t$ such that $S(t)$ exceeds a pre-determined threshold $ {b}$:
	\begin{align}\label{eq:tau}
		\tau( {b})=\inf\{t:S(t)\geq  {b}\}.
	\end{align}
	We evaluate the performance of a change detection scheme using the ARL to false alarm and the worst-case ADD (WADD) using Lorden's definitions \cite{lorden71} as follows:
	\begin{align*}
		\ARL(\tau) &= \E{\infty}[ \tau],\\
		\text{WADD}(\tau)& = \sup_{\nu\geq 1}~\esssup~\E{\nu}[(\tau-\nu+1)^+]{X_1,...,X_{\nu-1}}. 
	\end{align*}
	where $\sup$ is the supremum operator and and $\esssup$ is the essential supremum operator. The quickest change detection problem can be formulated as the following minimax problem\cite{lorden71}:
	\begin{align}\label{minimax}
		\begin{aligned}
			\min_\tau &\ \text{WADD}(\tau), \\
			\text{subject to} &\ \ARL(\tau)\geq \gamma,
		\end{aligned}
	\end{align}
	for some given $\gamma > 0$. We refer the interested reader to Chapter $6$ of \cite{poor2009quickest} for a comprehensive treatment and overview of Lorden's formulation of the quickest change detection problem.
	
	\section{Test statistic based on binning}\label{sec:algorithm}
	In this section, we derive a test statistic that can be recursively updated. If the post-change distribution belongs to a finite set of distributions, the GLRT statistic is the maximum of a finite set of CuSum statistics, one corresponding to each possible post-change distribution, and therefore has a recursion. To the best of our knowledge, the only other recursive test known in the literature for the case where the post-change distribution is not completely specified and does not belong to a finite set of distributions is proposed in \cite{lorden2008sequential}. The method in \cite{lorden2008sequential} requires that both the pre- and post-change distributions belong to a single parameter exponential family.
	
	Suppose that the post-change distribution $g\in D(f,N)$ and we observe the sequence $X_1=x_1, X_2=x_2, \ldots$ . If $g_N$ and ${f}_N$ are both known, Page's CuSum test statistic \cite{page54} for the binned observations is 
	\begin{align}
		S(t)&= \max_{1\leq k\leq t+1}\sum_{j=k}^{t} \log \frac{ g_N(x_j)}{ {f}_N(x_j)}. \label{eq:Sstat}
	\end{align}
	Note that $S(t)$ in \eqref{eq:Sstat} takes the value $0$ if $k=t+1$ is the maximizer. The test statistic $S(t)$ has a convenient recursion $S(t+1)=\max\{S(t)+\log(g_N(x_{t+1})/{f}_N(x_{t+1})),0\}$.
	
	In our problem formulation, $g_N$ is unknown. We thus replace $g_N(x_i)$ in \eqref{eq:Sstat} with its maximum likelihood estimator
	\begin{align*}
		g^{k:t}_N(x_i)=\frac{|\{x_r:k\leq r\leq t \text{ and $x_r\in I^N_j$}\}|}{t-k+1},		
	\end{align*}
	where $j$ is the unique integer such that $x_i\in I^N_j$. Note that in computing $g^{k:t}_N(x_i)$, we use only the samples $x_k,\ldots,x_t$. We then have the test statistic
	\begin{align}
		\max_{1\leq k\leq t+1}\sum_{i=k}^{t} \log \frac{ g^{k:t}_N(x_i)}{ {f}_N(x_i)}.\label{muMLE}
	\end{align}
	In the case where $t-k+1$ is small, the maximum likelihood estimator $g^{k:t}_N$  tends to over-fit the observed data and evaluating the likelihood of $x_i$ using $g^{k:t}_N(x_i)$ biases the test statistic. Furthermore, this couples the estimation of $g_N$ and the instantaneous likelihood ratio $\frac{ g_N(x_i)}{ {f}_N(x_i)}$. In order to compensate for this over-fitting, we choose not to include observations $x_i,...,x_t$ in the estimation of $g_N$. This also decouples the estimation of $g_N$ and the likelihood ratio $\frac{ g_N(x_i)}{ {f}_N(x_i)}$. However, if $x_i$ is the first observation occurring in the set $I^N_j$, we have $g^{k:i-1}_N(x_i)=0$. To avoid this, we define the regularized version of $g^{k:i-1}_N$ as
	\begin{align}\label{eqn:regularized}
		\widehat{g}^{k:i-1}_N(x_i)=
		\begin{cases}
			\frac{|\{x_r : k\leq r\leq i-1 \text{ and $x_r\in I^N_j$}\}|+R}{(N+H)R+i-k} &\text{if $k\leq i-1$,}\\
			f_N(x_i) &\text{otherwise,}
		\end{cases}
	\end{align}
	where $R$ is a fixed positive constant, and $j$ is the unique integer such that $x_i\in I^N_j$.  Our test statistic then becomes
	\begin{align}\label{eqn:teststatistics}
		\widehat{S}_N(t)&=\max_{1\leq k\leq t+1}\sum_{i=k}^t\log\frac{ \widehat{g}^{k:i-1}_N(x_i)}{f_N(x_i)},
	\end{align}
	with the stopping time
	\begin{align}
		\widehat{\tau}(b)&=\inf\{t:\widehat{S}_N(t)>b\}.\label{test_hS}
	\end{align}
	
	In practice, $R$ is chosen to be of the order of $N$ so that $\widehat{g}^{k:i-1}_N(x)$ approaches $1/N$ as $N\to\infty$. This controls the variability of \eqref{eqn:teststatistics} by controlling the range of values that $\log\frac{\widehat{g}^{k:i-1}_N(x_i)}{{f}_N(x_i)}$ can take. Computation of the test statistic \eqref{eqn:teststatistics} is inefficient as the estimator $\widehat{g}^{k:i-1}_N$ needs to be recomputed each time a new observation $X_t=x_t$ is made, leading to computational complexity increasing linearly w.r.t.~$t$. One way to prevent this increase in computational complexity is by searching for a change point from the previous most likely change point rather than from $t=1$, and also using observations from the previous most likely change point to the current observation to update the estimator for $g$. Our proposed BG-CuSum test statistic $\tS$ and test $\widetilde{\tau}$ are defined as follows: For each $t \geq 1$,
	\begin{align}
		\tS(t)& =\max_{\lambda_{t-1}\leq k\leq t+1}\sum_{i=k  }^t\log\frac{\widehat{g}^{\lambda_{t-1}:i-1}_N(x_i)}{{f}_N(x_i)},\label{BG-CuSum_S}\\
		\lambda_{t}& =
		\max\left\{\displaystyle\argmax_{\substack{\lambda_{t-1}\leq k\leq t+1\\k\neq \lambda_{t-1}+1}}\sum_{i=k}^{t}\log\frac{\widehat{g}^{\lambda_{t-1}:i-1}_N(x_i)}{{f}_N(x_i)}\right\},\label{def:lambda_t}\\
		\widetilde{\tau}(b)& =\inf\{t:\tS(t)\geq {b}\}, \label{BG-CuSum_test}
	\end{align}
	where $\lambda_0=1$, and $b$ is a fixed threshold. The outer maximum in \eqref{def:lambda_t} is to ensure that $\lambda_t$ is uniquely defined when there is more than one maximizer. Due to the design of the estimator \eqref{eqn:regularized}, we have $\widehat{g}^{\lambda_{t-1}:\lambda_{t-1}-1}_N(x_{\lambda_{t-1}})={f}_N(x_{\lambda_{t-1}})$ and
	\begin{align*}
		\sum_{i=\lambda_{t-1}  }^t\log\frac{\widehat{g}^{\lambda_{t-1}:i-1}_N(x_i)}{{f}_N(x_i)}=\sum_{i=\lambda_{t-1}+1  }^t\log\frac{\widehat{g}^{\lambda_{t-1}:i-1}_N(x_i)}{{f}_N(x_i)}.
	\end{align*}
	Thus, if $k=\lambda_{t-1}+1$ maximizes the sum $\sum_{i=k  }^t\log\frac{\widehat{g}^{\lambda_{t-1}:i-1}_N(x_i)}{{f}_N(x_i)}$, $k=\lambda_{t-1}$ also maximizes it. For this case, we choose the most likely change-point to be $\lambda_{t-1}$ rather than $\lambda_{t-1}+1$, as defined in \eqref{def:lambda_t}. Note also that if $\lambda_{t-1}=t$, then we have $\tS(t)=\tS(t-1)=0$, i.e., it takes more than one sample for our test statistic to move away from the zero boundary once it hits it. This is due to the way we define our estimator $\widehat{g}^{\lambda_{t-1}:t-1}_N$ in \eqref{eqn:regularized}, which at time $t$ utilizes samples starting from the last most likely change-point $\lambda_{t-1}=t$ to the previous time $t-1$ to estimate the post-change distribution, i.e., it simply uses $f_N$ as the estimator. From \eqref{def:lambda_t}, we then have $\lambda_t = t$. Furthermore, if $\tS(t-1)=0$ while $\tS(t) > 0$, then we must have $\lambda_{t-1}=t-1$.
	
	The test statistic $\tS(t)$ can be efficiently computed using a recursive update as shown in the following result.
	
	\begin{Theorem}\label{thm:Supdate}
		For each $t\geq0$, we have the update formula
		\begin{align}
			\tS(t+1)&=\max\left\{\tS(t)+\log\frac{\widehat{g}^{\lambda_t:t}_N(x_{t+1})}{{f}_N(x_{t+1})},0\right\},\label{Supdate}\\
			\lambda_{t+1}&=
			\begin{cases}
				\lambda_t &\text{if $\tS(t)+\log\frac{\widehat{g}^{\lambda_t:t}_N(x_{t+1})}{{f}_N(x_{t+1})}> 0$ or $\lambda_t=t+1$},\\
				t+2 &\text{otherwise},
			\end{cases}\label{lambdaupdate}
		\end{align}
		where $\tS(0)=0$ and $\lambda  _0=1$.
	\end{Theorem}
	\begin{IEEEproof}
		See Appendix \ref{sec:AppThm1}.
	\end{IEEEproof}
	
	Similar to the CuSum test, the renewal property of the test statistic and the fact that it is non-negative implies that the worst case change-point $\nu$ for the ADD is at $\nu=0$. We compare the performance of the stopping time \eqref{test_hS} with that of \eqref{BG-CuSum_test} using simulations in Section~\ref{sec:results}.
	
	\section{Properties of the BG-CuSum statistic}\label{sec:asymptotics}
	
	In this section, we present some properties of the BG-CuSum statistic in order to give insights into the asymptotic behavior of the ARL and ADD of our test. The proofs for the results in this section are provided in Appendix \ref{sec:AppThm3}. 
	
	\subsection{Estimating a lower bound for the ARL}\label{subsect:ARL_bound}
	
	In applications, a practitioner is required to set a threshold $b$ for the problem of interest. Thus, it is of practical interest to have an estimate of the ARL of our test w.r.t.\ the threshold $b$.  This is even more important in our context as it is not possible to set the threshold $b$ w.r.t.\ the WADD since it varies with the unknown post-change distribution. In this subsection, we derive a lower bound for the ARL of the BG-CuSum test.
	
	Define the stopping time 
	\begin{align*}
		\zeta( {b})=\inf\{t:\tS(t)\geq  {b}\text{ or }(\tS(t)\leq 0\text{ and } t> 2) \}.
	\end{align*}
	Note that the condition $t>2$ is used due to the lag in our test statistic; see the discussion after \eqref{BG-CuSum_test}. We have the following lower bound for the ARL of the BG-CuSum test.
	\begin{Proposition}\label{prop:ARL}
		For any threshold $b>0$, we have 
		\begin{align}
			\ARL(\widetilde{\tau}( {b}))&=\frac{\E{\infty}[\zeta( {b})]}{\P{\infty}(\tS(\zeta( {b}))\geq {b})}\geq e^{ {b} }. \label{ARL}
		\end{align}	
	\end{Proposition}

	\subsection{Growth rate of the BG-CuSum statistic}\label{subsect:ADD}
	
	We study the error bounds of the growth rate of the BG-CuSum statistic $\tS$ under the assumption that the observed samples $X_1=x_1,X_2=x_2,\ldots$ are generated by the post-change distribution $g$. We show that the growth rate $\tS(t)/t$ converges to the Kullback-Leibler divergence $\KLD{g_N}{{f}_N}$ $r$-quickly (see Section 2.4.3 of \cite{tartakovsky2014sequential}) for $r=1$ under $\P{1}$, which implies almost sure convergence. We then provide some heuristic insights into the asymptotic trade-off between the ARL and ADD of our BG-CuSum test as the threshold $b\to\infty$ in \eqref{BG-CuSum_test}.
	
	We first show that the regularized sample mean $\widehat{g}^{1:i}_N$ in \eqref{eqn:regularized} is close to $g_N$ with high probability when the sample size is large.
	
	\begin{Proposition}\label{prop:errorbound_for_mu}
		For any $\epsilon \in (0,1)$ and $x\in\mathbb{R}$, there exists $t_1\in\mathbb{N}$ such that for all $i\geq t_1$, we have
		\begin{align*}
			\P{1}(\left|\widehat{g}_N^{1:i}(x)-g_N(x)\right|\geq \epsilon) &\leq 2e^{-i\epsilon^2/2}.
		\end{align*}
	\end{Proposition}
	We next show that the instantaneous log-likelihood ratio $\log\frac{\widehat{g}_N^{1:i}(x)}{{f}_N(x)}$ is close to the true log-likelihood ratio $\log\frac{g_N(x)}{{f}_N(x)}$, in the sense that the probability of a deviation of $\epsilon$ decreases to zero exponentially as the number of samples increases.
	\begin{Proposition}\label{prop:error_instant_statistic}
		For any $\epsilon \in (0,1)$, there exists a $t_2\in\mathbb{N}$ such that for all $i\geq t_2$, we have
		\begin{align*}
			\P{1}(\left|\log\frac{\widehat{g}_N^{1:i}(X_{i+1})}{{f}_N(X_{i+1})}-\log\frac{g_N(X_{i+1})}{{f}_N(X_{i+1})}\right|\geq \epsilon) \leq2e^{-i(g_N^{\min}\epsilon)^2/8}
		\end{align*}
		where $g_N^{\min}=\min_x g_N(x)$.
	\end{Proposition}
	
	Putting Propositions \ref{prop:errorbound_for_mu} and \ref{prop:error_instant_statistic} together, we obtain the following theorem.
	
	\begin{Proposition}\label{thm:rate_of_growth}
		The empirical average $\frac{1}{t}\sum_{i=1}^t\log\frac{\widehat{g}_N^{1:i-1}(X_{i})}{{f}_N(X_{i})}$ converges to $\KLD{g_N}{{f}_N}$ $r$-quickly for $r=1$ under the distribution $\P_{1}$ as $t\to\infty$.
	\end{Proposition}
	
	From Proposition \ref{thm:rate_of_growth}, the probability of the growth rate of the BG-CuSum statistic $\tS(t)/t$ deviating from $\KLD{g_N}{{f}_N}$ by more than $\epsilon$ can be made arbitrarily small by increasing the number of samples $t$ after the change point. Heuristically, this means that the ADD increases linearly at a rate of $1/\KLD{g_N}{{f}_N}$ w.r.t. the threshold $b$. Unfortunately, due to technical difficulties introduced by having to estimate the post-change distribution using $\widehat{g}_N^{1:i-1}$, we are unable to quantify the asymptotic trade-off between ARL and ADD for our BG-CuSum test $\widetilde{\tau}$. In the following, we consider the asymptotic trade-off of the test $\widehat{\tau}$ in \eqref{test_hS} as an approximation of $\widetilde{\tau}$, and use simulation results in Section~\ref{subsec: ADD_asymptotic} to verify that a similar trade-off applies for the BG-CuSum test.
	
	\begin{Proposition}\label{thm:add_arl_shat}
		The stopping time $\widehat{\tau}(b)$ defined in \eqref{test_hS} satisfies $$\ARL(\widehat{\tau}(b))\geq e^b,$$ and $$\text{WADD}(\widehat{\tau}(b))\leq\frac{b}{\KLD{g_N}{f_N}} + o(b)\quad\text{as $b\to \infty$}.$$ Furthermore if both $f$ and $g$ are discrete distributions (i.e., $p_0=q_0=0$), the stopping time $\widehat{\tau}(b)$ is asymptotically optimal.
	\end{Proposition}

	\section{Numerical Results}
	\label{sec:results}
	In this section, we first compare the performance of our proposed BG-CuSum test with two other non-parametric change detection methods in the literature. We first perform simulations, and then we verify the performance of our method on real activity tracking data from \cite{kwapisz2011activity}.
	
	\subsection{Synthetic data}
	
	In our first set of simulations, we set the parameters $H=0$, $N=16$, $R=16$ and choose the pre-change distribution to be the standard normal distribution $\calN(0,1)$. All the post-change distributions we use in our simulations are in $D({f},N)$ for this choice of parameters.
	
	In our first experiment, we compare the performance of our method with \cite{hawkins10}, in which it is assumed that the change is an unknown shift in the location parameter. We also compare the performance of our method with \cite{darkhovskii1988nonparametric}, in which it is assumed that the means of the pre- and post-change distributions are different, and both distributions are unknown. In our simulation, we let the post-change distribution be $\calN(\delta,1)$. We control the ARL at $500$ and set the change-point $\nu=300$ for both methods while varying $\delta$. The average detection delay is computed from $50,000$ Monte Carlo trials and shown in Table \ref{tab:mean}, where the smallest ADDs for each ARL are highlighted in boldface. We see that our method, despite not assuming that the change is a mean shift, achieves a comparable ADD as the method in \cite{hawkins10}. Furthermore, we see that our method outperforms the method in \cite{darkhovskii1988nonparametric}.
	\begin{table}[!htb]
		\centering
		\caption{ADD for post-change distribution $\calN(\delta,1)$ with $\ARL=500$.}\label{tab:mean}
		\begin{tabular}{|c|c|c|c|c|c|}
			\hline
			$\delta$ & 0.125 &0.75 & 1.5 & 2.25 & 3    \\
			\hline\hline
			Hawkins\cite{hawkins10} & 428.60 & 18.1 & \textbf{6.6} & 4.5 & 3.9  \\
			\hline
			\shortstack{Darkhovskii\cite{darkhovskii1988nonparametric}\\($N=280, {b}=0.25,c=0.1$)} & 458.71 & 26.15 & 13.02 & 8.55 & 6.58  \\[0.5ex]
			\hline
			BG-CuSum & \textbf{344.78} & \textbf{17.9} & \textbf{6.6} & \textbf{3.2} & \textbf{2.3} \\
			\hline
		\end{tabular}
	\end{table}
	
	We next consider a shift in variance for the post-change distribution. The method in \cite{hawkins10}, for example, will not be able to detect this change accurately as it assumes that the change in distribution is a shift in mean. Therefore, we also compare our method with the KS-CPM method\cite{ross12}, which is a non-parametric test that makes use of the Kolmogorov-Smirnov statistic to construct a sequential 2-sample test to test for a change-point. We control the ARL at $500$ and change-point $\nu=300$ for all methods while varying $\delta$ for the post-change distribution $g\sim\calN(0,\delta^2)$. The average detection delays computed from $50,000$ Monte Carlo trials are shown in Table \ref{tab:variance}. We see that our method outperforms both \cite{hawkins10} and \cite{ross12} in the ADD.
	%<*tag:r1c3>
	We note that $\KLD{g_N}{f_N}$ decreases as $\delta$ varies from $0.2$ to $1$, and increases as $\delta$ varies from $1$ to $2$. Thus, the observation that the ADD increases as $\delta$ varies from $0.2$ to $1$, and decreases as $\delta$ varies from $1$ to $2$ agrees with our expectation.
	%</tag:r1c3>
	\begin{table}[!htb]
		\centering
		\caption{ADD for post-change distribution $\calN(0,\delta^2)$ with $\ARL=500$.}\label{tab:variance}
		\begin{tabular}{|c|c|c|c|c|c|}
			\hline
			$\delta$ &0.2 & 0.33 &0.5 & 1.5 & 2     \\
			\hline\hline
			Hawkins\cite{hawkins10} & 361.3& 391.5 & 438.5 & 149.6 & 75.3  \\
			\hline
			KS-CPM\cite{ross12} & 27.2& 37.2 & 84.7 & 140.6 & 49.2  \\
			\hline
			BG-CuSum &\textbf{10.5}& \textbf{17.4} & \textbf{33.3} & \textbf{45.2} & \textbf{21.5} \\
			\hline
		\end{tabular}
	\end{table}
	
	Next, we test our method with the Laplace post-change distribution with the probability density function $g(x) = \frac{1}{2\times 0.7071}e^{-\frac{{|x|}}{0.7071}}$. The location and scale parameter are chosen such that the first and second order moments of $g$ and ${f}$ are equal. We set the $\ARL=500$ and two different values for the change-point $\nu$. We performed $50,000$ Monte Carlo trials to obtain Table \ref{tab:laplace}. The results show the method is able to identify the change from a normal to a Laplace distribution with the smallest ADD out of the three methods studied.
	
	\begin{table}[!htb]
		\centering
		\caption {ADD for Laplace post-change distribution with $\ARL=500$.} \label{tab:laplace}
		\begin{tabular}{|c|c|c|}
			\hline
			$\nu$& 50 & 300\\
			\hline\hline
			Hawkins\cite{hawkins10}& 592 & 828\\
			\hline
			KS-CPM\cite{ross12} & 284 & 217 \\
			\hline		
			BG-CuSum & \textbf{156} & \textbf{154}  \\
			\hline
		\end{tabular}
	\end{table}

	\begin{figure}
		\centering
		\centerline{\includegraphics[width=10.0cm]{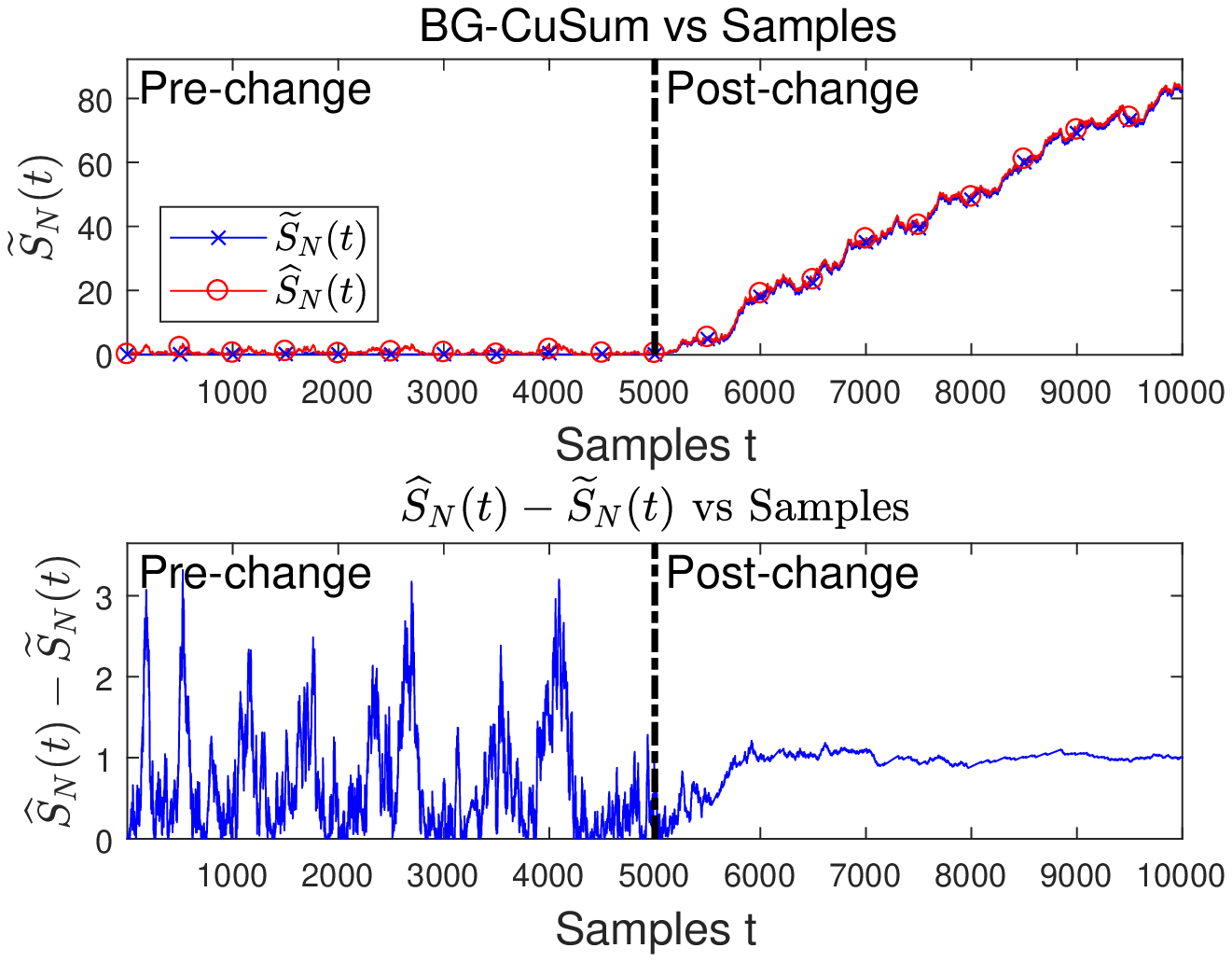}}
		%  \vspace{1.5cm}
		\caption{Examples of test statistics $\widehat{S}(t)$, $\widetilde{S}(t)$ and $\widehat{S}(t)-\widetilde{S}(t)$ as a function of $t$ for $\KLD{g_N}{f_N}=2$.}
		\label{fig:compare_bg_stat_vs_opt_stat_continuous}
	\end{figure}
	\begin{figure}
		\centering
		\centerline{\includegraphics[width=10.0cm]{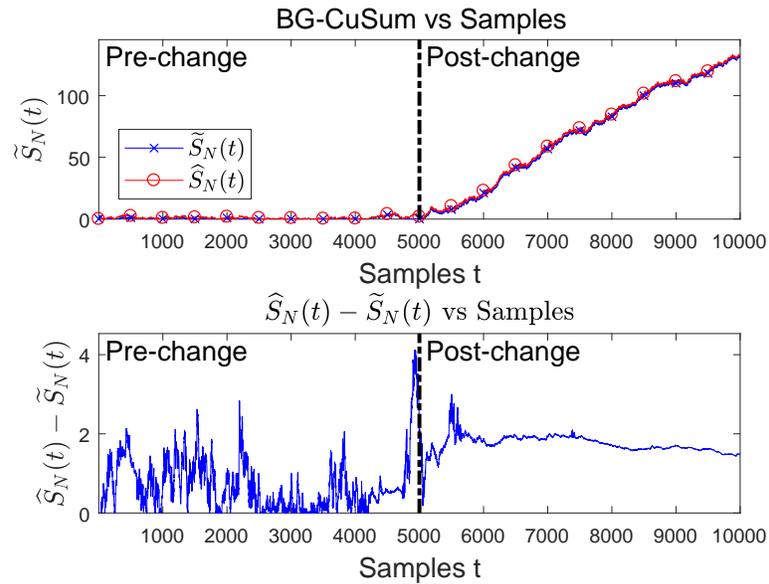}}
		%  \vspace{1.5cm}
		\caption{Examples of test statistics $\widehat{S}(t)$, $\widetilde{S}(t)$ and $\widehat{S}(t)-\widetilde{S}(t)$ as a function of $t$ when the change is in the discrete component such that the pre-change distribution is $0.5\calN(0,1)+0.25\delta_{-1}+0.25\delta_{1}$ and post-change distribution is $0.5\calN(0,1)+0.33\delta_{-1}+0.17\delta_{1}$.}
		\label{fig:compare_bg_stat_vs_opt_stat_discrete}
	\end{figure}
	
	In \figref{fig:compare_bg_stat_vs_opt_stat_continuous} and \figref{fig:compare_bg_stat_vs_opt_stat_discrete}, we show how the BG-CuSum test statistic $\tS(t)$ behaves for two different post-change distributions. In  \figref{fig:compare_bg_stat_vs_opt_stat_continuous}, the pre-change distribution is $\calN(0,1)$ and the post-change distribution is $\calN(0.2,1)$. In  \figref{fig:compare_bg_stat_vs_opt_stat_discrete}, we consider distributions with both discrete and absolutely continuous components. The change is in the discrete component where the pre-change distribution is $0.5\calN(0,1)+0.25\delta_{-1}+0.25\delta_{1}$ and the post-change distribution is $0.5\calN(0,1)+0.33\delta_{-1}+0.17\delta_{1}$. We observe, in both cases, that $\tS(t)$ remains low during the pre-change regime and quickly rises in the post-change regime in both cases. Furthermore, the proposed recursive test statistics $\tS$ is observed to track the test statistic $\widehat{S}_N$, which has known asymptotic properties as seen in Proposition~\ref{thm:add_arl_shat}.
	\subsection{Choice of \texorpdfstring{$N$}{N}}
	
	\begin{figure}[!htb]
		\centering
		\centerline{\includegraphics[width=10.0cm]{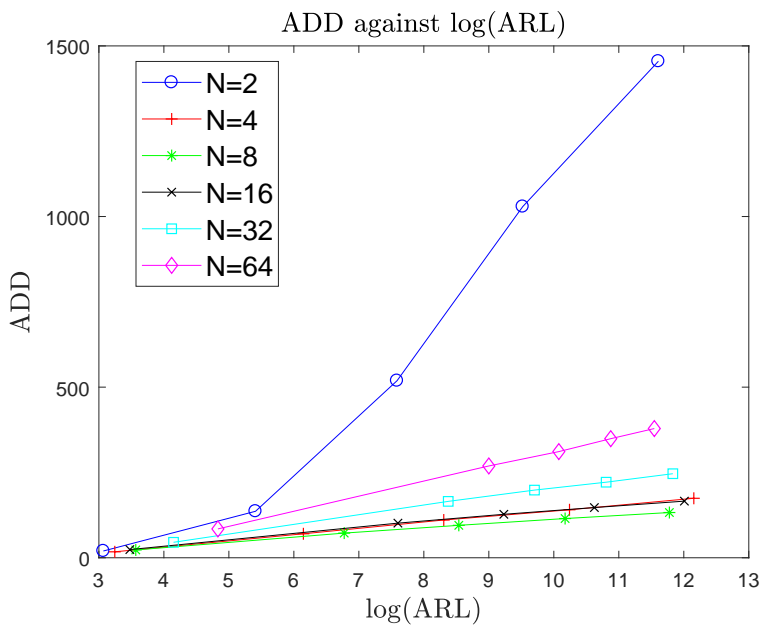}}
		%  \vspace{1.5cm}
		\caption{Graph of ADD against $\log(\text{ARL})$ for varying values of $N$ with ${f}=\calN(0,1)$ and $g=0.6\calN(1,1)+0.4\calN(-1,1)$.}
		\label{fig:Performance_vs_N}
	\end{figure}
	
	Before applying the BG-CuSum test, the user has to choose an appropriate number of bins $N$. In the supplementary material \cite{supplementary}, we have provided a procedure for determining a suitable choice of $N$. In this subsection, we present several results to illustrate the guiding principles for choosing $N$. First, we compare the ARL-ADD performance of BG-CuSum using different values of $N$. We performed the experiments with $\calN(0,1)$ as the pre-change distribution, and Gaussian mixture model $0.6\calN(1,1)+0.4\calN(-1,1)$ as the post-change distribution for $N=2,4,8,16,32,64,$ using $5000$ Monte Carlo trials to obtain the ADD and ARL. The KL divergence $\KLD{g_N}{f_N}$ for respective values of $N$ are $0.0094,0.730,0.1164,0.1420,0.1565$ and $0.1645$. From \figref{fig:Performance_vs_N}, we observe that the performance of BG-CuSum improves as $N$ increases from $2$ to $8$ and degrades as $N$ increases from $8$ to $64$. One reason for this is that the benefit of having a larger $\KLD{g_N}{f_N}$ does not out-weigh the larger number of samples required to accurately estimate the unknown post-change distribution $g_N$ when $N$ is large.

	\subsection{Asymptotic behaviour of the ARL and ADD of \texorpdfstring{$\widetilde{\tau}$}{tau}}\label{subsec: ADD_asymptotic}
	
\begin{figure}
		\centering
		\centerline{\includegraphics[width=10.0cm]{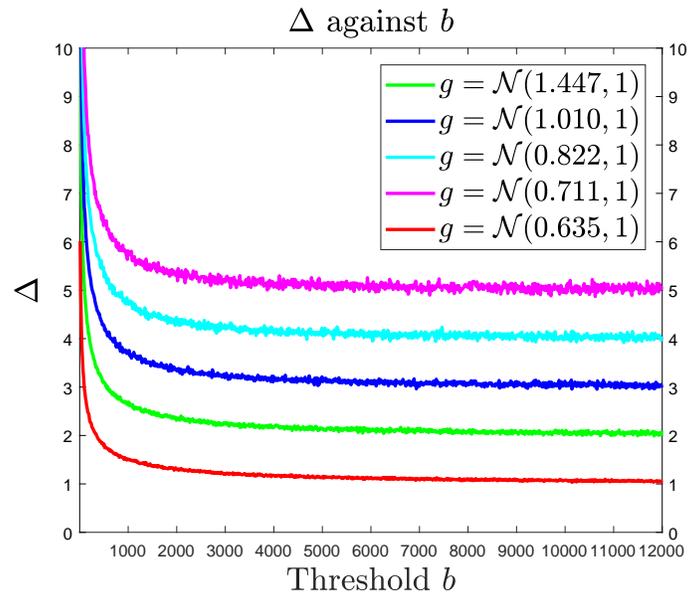}}
		%  \vspace{1.5cm}
		\caption{Plot of $\Delta(b)$ against threshold $ {b}$.}
		\label{fig:ADDvsGamma}
	\end{figure}

	In Section~\ref{sec:asymptotics}, we showed that the growth of the BG-CuSum test statistic can be made arbitrarily close to $\KLD{g_N}{{f}_N}$ with a sufficiently large number of samples $t$. Heuristically, from Proposition~\ref{thm:add_arl_shat}, using $\widehat{S}(t)$ in \eqref{eqn:teststatistics} as an approximation of the BG-CuSum test statistic $\tS(t)$, this implies that the average detection delay would grow at a rate of $1/\KLD{g_N}{{f}_N}$. To demonstrate this, we let $N=64$, $\calN(0,1)$ as the pre-change distribution, and $g$ to be one of several normal distributions with different means and variance 1 so that we have $\KLD{g_N}{{f}_N}=1,\frac{1}{2},\frac{1}{3},\frac{1}{4},\frac{1}{5}$, respectively. Therefore we expect the asymptotic gradient of the ADD w.r.t.\ $ {b}$ to be $1,2,3,4$ and $5$ respectively. We performed $1000$ Monte Carlo trials to estimate the ADD for different values of $ {b}$. \figref{fig:ADDvsGamma} shows the plot of ADD against $b$ and 
	\begin{align*}
		\Delta( {b})=\frac{\text{ADD}(\tau( {b}+h))-\text{ADD}(\tau( {b}))}{h},	
	\end{align*}		
	which approximates the gradient of WADD w.r.t.\ $b$. We used a step-size of $h=12$ to generate \figref{fig:ADDvsGamma}. We see that $\Delta(b)$ tends to $1,2,3,4$ and $5$ respectively as $b$ tends to infinity, which agrees with those predicted by our heuristic.
	
\begin{figure}
		\centering
		\centerline{\includegraphics[width=10.0cm]{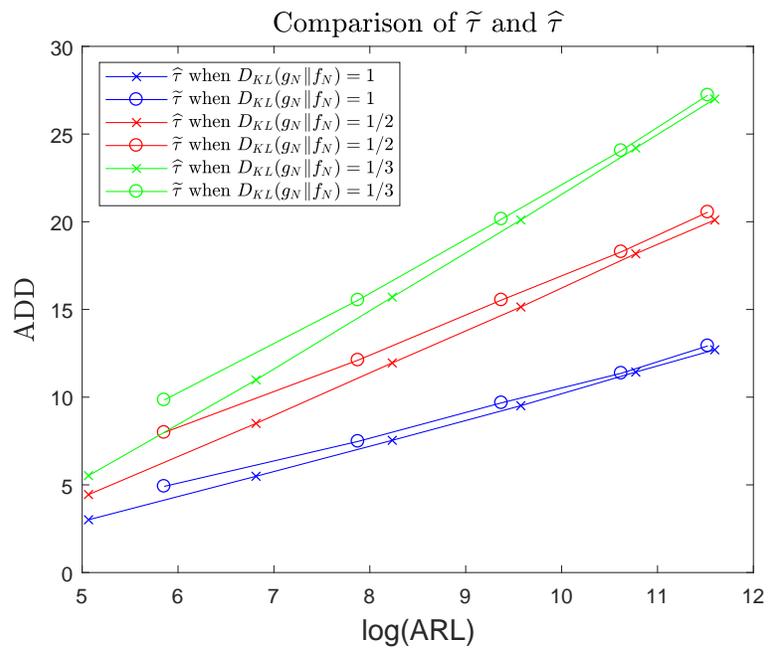}}
		%  \vspace{1.5cm}
		\caption{Comparison of the trade-off between ADD and $\log(\text{ARL})$ for $\widetilde{\tau}$ and $\widehat{\tau}$.}
		\label{fig:tauhatvstautilde}
	\end{figure}

	Next, we compare the asymptotic performance of $\widetilde{\tau}$ in \eqref{BG-CuSum_test} and $\widehat{\tau}$ in \eqref{test_hS}. We perform simulations with $N=4$ and $f=\mathcal{N}(0,1)$ and $g$ to be one of several normal distributions with different means and variance $1$ so that we have $\KLD{g_N}{f_N}=1,2\text{ and }3$. We perform $5000$ Monte Carlo trials to estimate the ARL and ADD of both $\widehat{\tau}$ and $\widetilde{\tau}$. \figref{fig:tauhatvstautilde} shows that the ADD of $\widetilde{\tau}$ remains close to $\widehat{\tau}$ as the ARL becomes large.

	\subsection{WISDM Actitracker Dataset}

	\begin{figure}[!htb]
		\begin{minipage}[b]{.99\linewidth}
			\centering
			\centerline{\includegraphics[width=10.0cm]{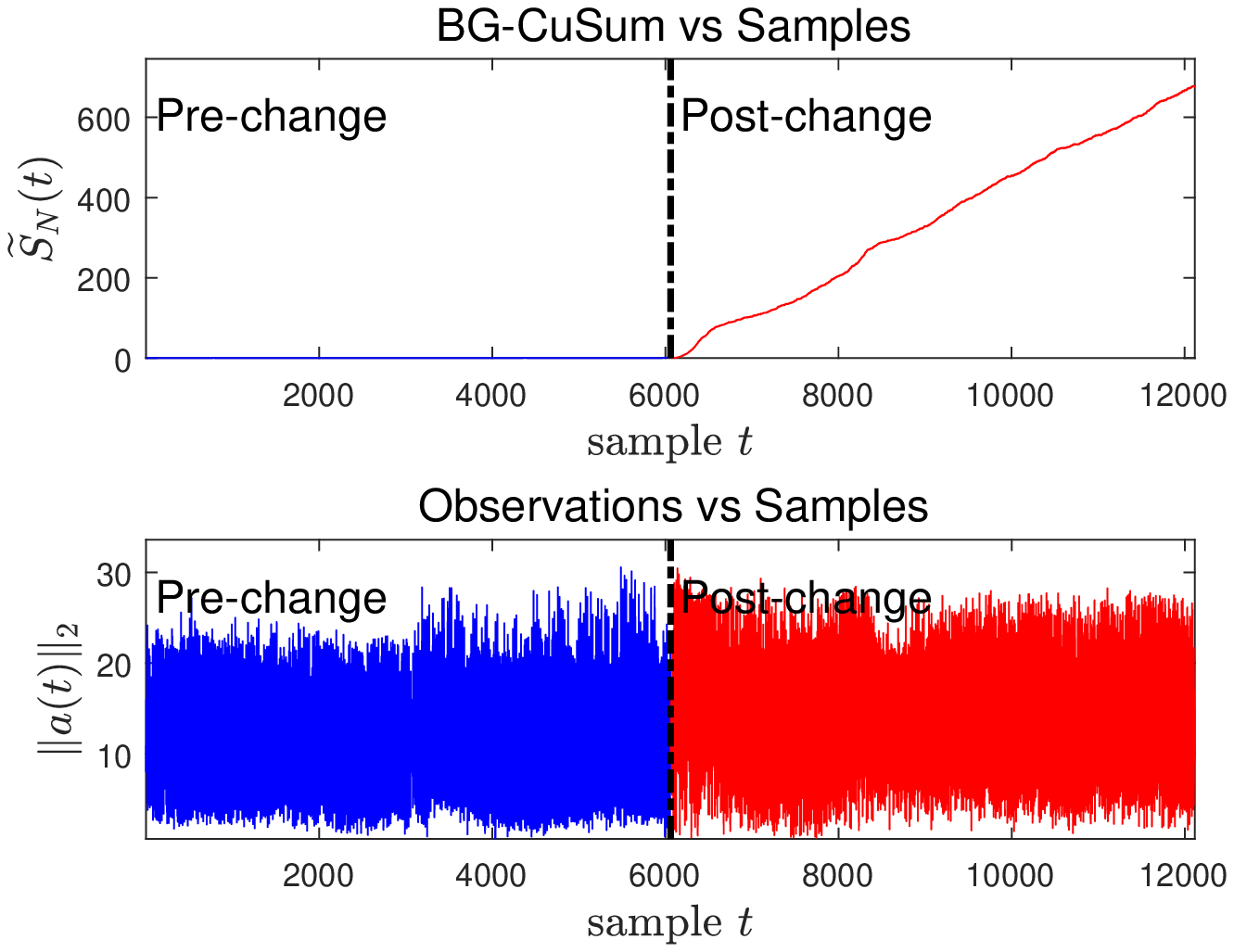}}
			%  \vspace{1.5cm}
			\centerline{(a)}\medskip
		\end{minipage}
		\hfill
		\begin{minipage}[b]{0.99\linewidth}
			\centering
			\centerline{\includegraphics[width=10.0cm]{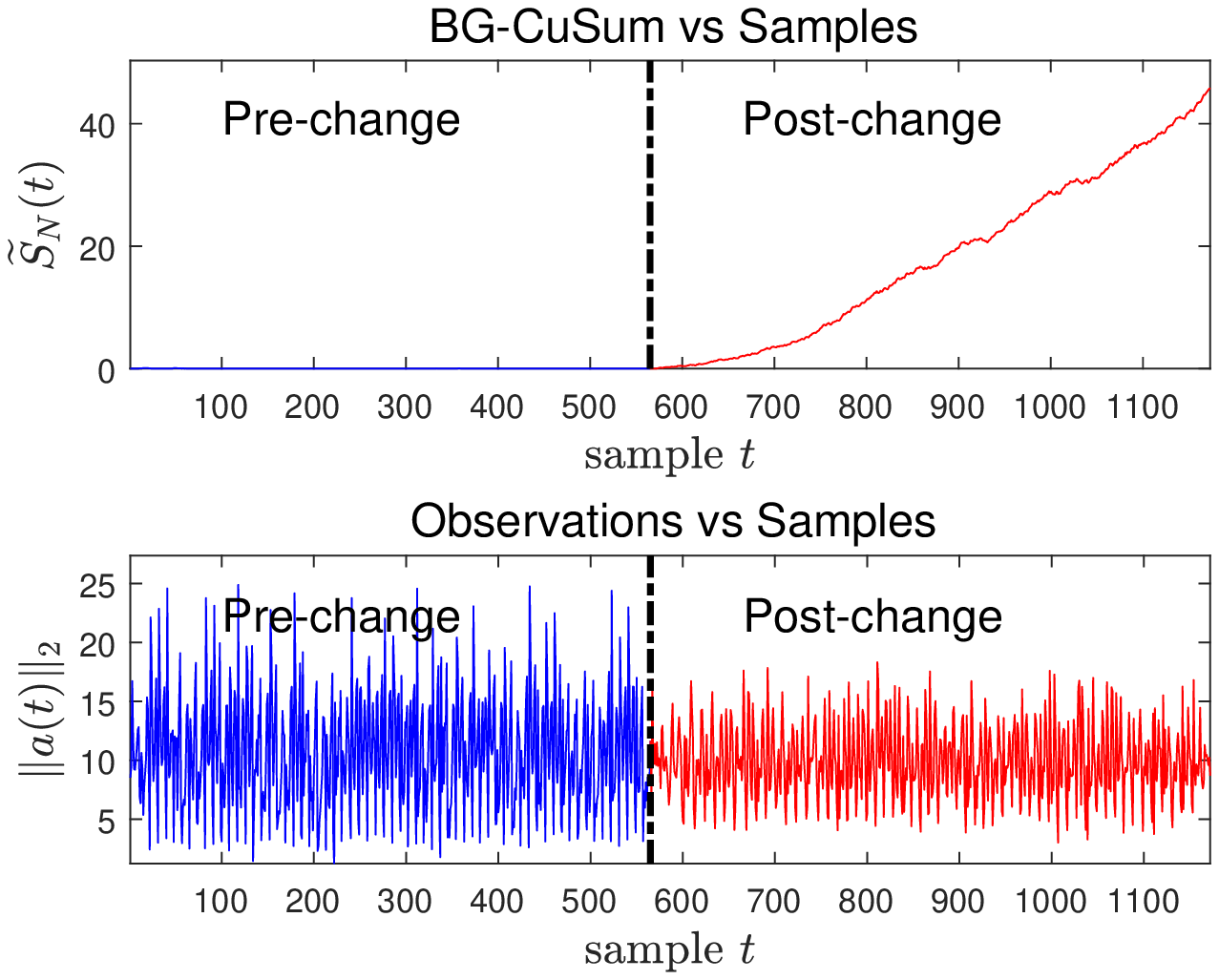}}
			%  \vspace{1.5cm}
			\centerline{(b)}\medskip
		\end{minipage}
		\caption{Examples of trial performed for pre-change activity of walking, and post-change activity of (a) jogging  and (b) ascending upstairs. The black dotted line indicates the boundary between the pre-change and post-change regimes. }
		\label{fig:real_data}
	\end{figure}
	
	\begin{figure}[!htb]
		\centering
		\centerline{\includegraphics[width=10.0cm]{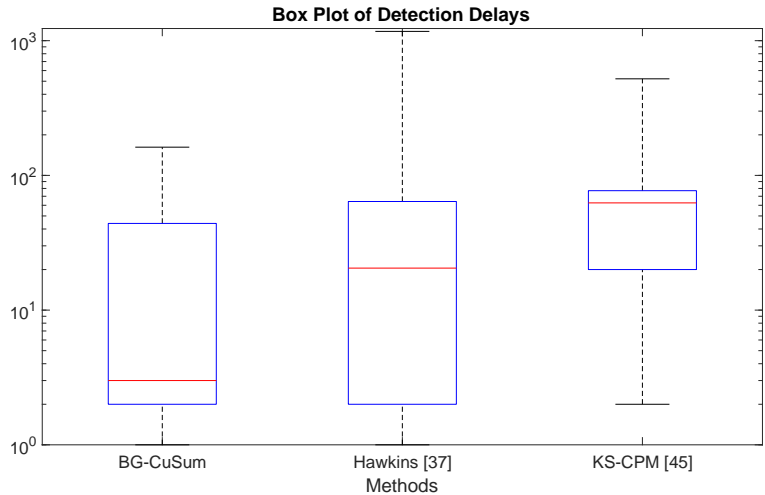}}
		\caption{Box-plot of the detection delays of each algorithm when the ARL is $6000$.}\label{fig:real_data_boxplot}
	\end{figure}
	
	We now apply our BG-CuSum test on real activity tracking data using the WISDM Actitracker dataset \cite{kwapisz2011activity}, which contains 1,098,207 samples from the accelerometers of Android phones. The dataset was collected from $29$ volunteer subjects who each carried an Android phone in their front leg pocket while performing a set of activities such as walking, jogging, ascending stairs, descending stairs, sitting and standing. The signals from the accelerometers were collected at $20$ samples per second. The movements generated when each activity by each individual can be assumed to be consistent across time. Hence we assume that the samples generated are i.i.d.\ for each activity.    
	
	To test the effectiveness of the BG-CuSum test on real data, we apply it to detect the changes in the subject's activity. We aim to detect the change from walking activity to any other activity. There are $45$ segments in the dataset in which there is a switch from walking to other activities. For each of these segments, we use the first half of the samples from the walking activity period to learn the pre-change pdf $f$ for the data. 
	
	As each sample at time $t$ from the phone's accelerometer is a 3-dimensional vector $\mathbf{a}(t)$, we perform change detection on the  sequence $x_t=\|\mathbf{a}(t)\|_2$ instead. We set the number of bins in the BG-CuSum test to be $H=0$ and $N=32$. Using the first $T$ samples from each walking activity segment, we estimate the boundaries of the intervals $I^N_j$, such that $\int_{I^N_j} f(x) \ \ud{x}=1/N$ by setting $I^N_1 =\left(-\infty,x_{\left(\floor{T/N}\right)}\right]$, $I^N_j =\left(x_{(\floor{(j-1)T/N})},x_{(\floor{jT/N})}\right]$ for $1<j<N-1$, and $I^N_{N} =\left(x_{(\floor{(N-1)T/N})},\infty\right)$, where $x_{(n)}$ is the $n$-th order statistic of $x_t$.
	
	In order to control the ARL of the BG-CuSum test to be $6000$, we set the threshold $ {b}$ to be $2.69$. The ADD for BG-CuSum on the WISDM Actitracker dataset is found to be about $30.2$.
	
	In \figref{fig:real_data}, we present some examples of the performance of BG-CuSum. Similar to the simulations, we observe that in both cases, the test statistic $\tS(t)$ remains low during the pre-change regime and quickly rises in the post-change regime. In \figref{fig:real_data_boxplot}, we present the notched boxplots of the detection delays of different algorithms when the ARL is set at $6000$. We observe that the BG-CuSum test out-performs both the KS-CPM\cite{ross12} and Hawkins\cite{hawkins10} algorithm.

	\section{Conclusion}\label{sec:conclude}
	
	We have studied the sequential change detection problem when the pre-change distribution $f$ is known, and the post-change distribution $g$ lies in a family of distributions with $k$-th moment differing from ${{f}}$ by at least $\epsilon$. We proposed a sequential change detection method that partitions the sample space into bins, and  a test statistic that can be updated recursively in real time with low complexity. We analyzed the growth rate of our test statistic and used it to heuristically deduce the asymptotic relationship between the ADD and the ARL. Tests on both synthetic and real data suggest that our proposed BG-CuSum test outperforms several other non-parametric tests in the literature. Furthermore, simulations indicate that the BG-CuSum test approaches the performance of $\widehat{\tau}$, which has known asymptotic properties, as the ARL becomes large. We provided a lower bound on the ARL of the BG-CuSum test to aid the setting of the threshold $b$. One direction for future work would be to derive the WADD for the BG-CuSum test. This remains an open research problem due to the technical difficulties introduced by having to estimate the post-change distribution using an estimated change-point.
	
	Although we have assumed that the pre- and post-change distributions are defined on $\Real$, the BG-CuSum test derived in this paper can be applied to cases where $f$ and $g$ are generalized pdfs on $\mathbb{R}^n$ with $n>1$. To see how this can be done, we assume that $g\in D({f},N)$, where $N={N_0}^n$ is a power of $n$. We can then divide $\mathbb{R}^n$ into $N$ equi-probable sets w.r.t. $f_c$ by sequentially dividing each dimension into $N_0$ equi-probable intervals. After obtaining these $N$ sets, we can apply the BG-CuSum test directly. The results derived in Section \ref{sec:asymptotics} can also be extended to the case where the distributions are on $\Real^n$. However, the amount of data required to learn the pre-change distribution ${f}$ increases quickly w.r.t.\ $n$, which limits its application in practice.
	
	%<*tag:r1c1>
	The assumption that the post change distribution $g$ is absolutely continuous w.r.t. $f$ can be further relaxed. For example, the results developed in Section~\ref{sec:algorithm} and~\ref{sec:asymptotics} extend to the case with a continuous pre-change distribution $f$ and a discrete post-change distribution $g$. The recursive update scheme in Theorem~\ref{thm:Supdate} and the lower bound on the ARL in Proposition~\ref{prop:ARL} still holds. However, as $g$ is not absolutely continuous w.r.t. $f$, the instantaneous empirical log-likelihood ratio $\log\frac{\widehat{g}_N^{1:i-1}(X_{i})}{{f}_N(X_{i})}=\infty$ and the instantaneous log-likelihood ratio $\log\frac{\widehat{g}_N(X_{i})}{{f}_N(X_{i})}=\infty$ with positive probability under $\P_{1}$. Thus, we have that $\KLD{g_N}{{f}_N}=\infty$. Furthermore, in Proposition~\ref{thm:rate_of_growth}, the empirical average $\frac{1}{t}\sum_{i=1}^t\log\frac{\widehat{g}_N^{1:i-1}(X_{i})}{{f}_N(X_{i})}$ diverges to infinity almost surely.
	%</tag:r1c1>
	
	One possible direction for future research is to extend the approach to quickest change detection in Hidden Markov Models\cite{fuh2003sprt,fuh2004asymptotic} when the post-change transition probability matrix is unknown. In order to compute the likelihood function, a similar estimation scheme for the transition probability matrix can be derived using the maximum-likelihood state estimates of the observed samples. However, more work is required to derive the asymptotic operating characteristics of the stopping time.	
	%\subsection{Other Error bounds}
	
	%\begin{Lemma}
	%Given $ \epsilon ,\delta>0$, exist a $t_6\in\mathbb{N}$ such that for all %$t-k\geq t_6$ and any $j\in\{1,...,N\}$ we have
	%  \begin{eqnarray*}
	%    %\P{}(\left|\frac{1}{t-k}\sum_{i=k+1}^t\log\frac{\widehat{g}_N^{1:i-1}(x_{i})}{{f}_N(%x_{i})}-\frac{1}{t-k}\sum_{i=k+1}^t\log\frac{g_N^{k:i-1}(x_{i})}{{f}_N(x_{%i})}\right|\geq \epsilon) \leq\delta.
	%  \end{eqnarray*}
	%\end{Lemma}
	
	%\begin{Lemma}
	%Given $ \epsilon ,\delta>0$, exist a $t_7\in\mathbb{N}$ such that for all %$t-k\geq t_7$ and any $j\in\{1,...,N\}$ we have
	%  \begin{eqnarray*}
	%    %\P{}(\left|D(g_N\|{f}_N)-\frac{1}{t-k}\sum_{i=k+1}^t\log\frac{g_N^{k:i-1%}(x_{i})}{{f}_N(x_{i})}\right|\geq \epsilon) \leq\delta.
	%  \end{eqnarray*}
	%\end{Lemma}
	\appendices
	\section{Proof of Theorem \ref{thm:Supdate}}\label{sec:AppThm1}
	In order to show that $\tS$ can be computed recursively, we require the following lemmas. %\red{[Please check proofs, have made various changes.]}
	\begin{Lemma_A}
		\label{lem:main}
		Suppose $\lambda_p=\lambda_{p+1}=p+1$ and $\tS(p+2),\ldots,\tS(p+n)>0$ for some $n\geq 1$, then we have
		$$\lambda_{p+2}=\lambda_{p+3}=\ldots=\lambda_{p+n}=p+1.$$
	\end{Lemma_A}
	\begin{IEEEproof}
		We prove by contradiction. Suppose that there exists $t\in\{2,...,n\}$ such that $\lambda_{p+t}\neq p+1$. Let $t_0$ be the smallest of all such indices $t$. 
		Since $\lambda_{p+t_0}\neq p+1$, following the definition of $\lambda_{p+t_0}$, we obtain $\lambda_{p+t_0}> \lambda_{p+t_0-1}=p+1$.
		Furthermore, by \eqref{def:lambda_t}, $\lambda_{p+t_0}\neq \lambda_{p+t_0-1}+1=p+2$. Thus we have $\lambda_{p+t_0}>p+2$. 
		From \eqref{BG-CuSum_S}, we have
		\begin{align}
			\tS(p+t_0)&=\max_{\lambda_{p+t_0-1}\leq k\leq p+t_0+1}\sum_{i=k}^{p+t_0}\log\frac{\widehat{g}^{\lambda_{p+t_0-1}:i-1}_N(x_i)}{{f}_N(x_i)}\nonumber\\
			&=\max_{p+1\leq k\leq p+t_0+1}\sum_{i=k}^{p+t_0}\log\frac{\widehat{g}^{p+1:i-1}_N(x_i)}{{f}_N(x_i)}\label{eqn:Spj0}\\
			&= \sum_{i=\lambda_{p+t_0}}^{p+t_0}\log\frac{\widehat{g}^{p+1:i-1}_N(x_i)}{{f}_N(x_i)},\nonumber
		\end{align}
		where the last equality follows from the definition \eqref{def:lambda_t}. We then obtain
		\begin{align}
			\sum_{i=p+1}^{\lambda_{p+t_0}-1}\log\frac{\widehat{g}^{p+1:i-1}_N(x_i)}{{f}_N(x_i)}=\sum_{i=p+1}^{p+t_0}\log\frac{\widehat{g}^{p+1:i-1}_N(x_i)}{{f}_N(x_i)}-\tS(p+t_0) \leq 0,\label{sum_lambda}
		\end{align}
		where the inequality follows from \eqref{eqn:Spj0}. Since $p+1<\lambda_{p+t_0}-1\leq p+t_0$, let $\lambda_{p+t_0}-1=p+t_1$ where $2\leq t_1 \leq t_0 \leq n$. Since $t_0$ is the smallest $t$ such that $\lambda_{p+t}\neq p+1$, we have $\lambda_{p+t_1-1}=p+1$ and
		\begin{align}
			\tS(p+t_1)&=\sum_{i=\lambda_{p+t_1}}^{p+t_1}\log\frac{\widehat{g}^{\lambda_{p+t_1-1}:i-1}_N(x_i)}{{f}_N(x_i)}\nonumber\\
			&=\sum_{i=\lambda_{p+t_1}}^{\lambda_{p+t_0}-1}\log\frac{\widehat{g}^{p+1:i-1}_N(x_i)}{{f}_N(x_i)}\leq 0,\label{eqn:lessthanzero}
		\end{align}
		where the last inequality follows trivially if $t_1= t_0$ and from \eqref{sum_lambda} if $t_1 < t_0$. This inequality contradicts our assumption that $\tS(p+2),\ldots,\tS(p+n)>0$. The lemma is now proved.\\
	\end{IEEEproof}
	\begin{Lemma_A}\label{lem:recursive}
		If $\tS(t+1)>0$, then $\tS(t+1)=\tS(t)+\log\frac{\widehat{g}^{\lambda_{t}:t}_N(x_{t+1})}{{f}_N(x_{t+1})}$
	\end{Lemma_A}
	\begin{IEEEproof}
		We first consider the case where $\tS(t)=0$. Since $\tS(t+1)>0$, we must have $\lambda_t=t$. We then obtain
		\begin{align*}
			\tS(t+1)&=\max_{\lambda_{t}\leq k\leq t+2}\sum_{i=k}^{t+1}\log\frac{\widehat{g}^{\lambda_{t}:i-1}_N(x_i)}{{f}_N(x_i)}\\
			&=\max_{t\leq k\leq t+2}\sum_{i=k}^{t+1}\log\frac{\widehat{g}^{t:i-1}_N(x_i)}{{f}_N(x_i)}\\
			&=\log\frac{\widehat{g}^{t:t}_N(x_{t+1})}{{f}_N(x_{t+1})}\\
			&=\tS(t)+\log\frac{\widehat{g}^{\lambda_t:t}_N(x_{t+1})}{{f}_N(x_{t+1})}.
		\end{align*}
		For the case where $\tS(t)>0$, let $n\geq 0$ be the largest integer such that $\tS(t-n),\ldots,\tS(t)>0$. If $n=0$, we trivially have $\lambda_{t-1}=\lambda_t=t-1$. If $n>0$, since $\tS(t-n-1)=0$, we have $\lambda_{t-n-1}=t-n-1$ and Lemma~\ref{lem:main} yields $\lambda_{t-n}=\ldots=\lambda_{t}=t-n-1$. We then obtain
		\begin{align*}
			\tS(t+1)
			&=\max_{\lambda_{t}\leq k\leq t+1}\sum_{i=k}^{t+1}\log\frac{\widehat{g}^{\lambda_{t}:i-1}_N(x_i)}{{f}_N(x_i)}\\
			&=\max_{\lambda_{t}\leq k\leq t+1}\left\{\sum_{i=k}^{t}\log\frac{\widehat{g}^{\lambda_{t}:i-1}_N(x_i)}{{f}_N(x_i)}+\log\frac{\widehat{g}^{\lambda_t:t}_N(x_{t+1})}{{f}_N(x_{t+1})}\right\}\\
			&=\max_{\lambda_{t-1}\leq k\leq t+1}\left\{\sum_{i=k}^{t}\log\frac{\widehat{g}^{\lambda_{t-1}:i-1}_N(x_i)}{{f}_N(x_i)}\right\}+\log\frac{\widehat{g}^{\lambda_t:t}_N(x_{t+1})}{{f}_N(x_{t+1})}\\
			&=\tS(t)+\log\frac{\widehat{g}^{\lambda_t:t}_N(x_{t+1})}{{f}_N(x_{t+1})},\\
		\end{align*}
		and the proof is complete.
	\end{IEEEproof}
	We are now ready to show that the statistic $\tS$ can be updated recursively. From Lemma \ref{lem:recursive} and noting that $\tS(t)\geq0$ for all $t$, \eqref{Supdate} follows immediately. If $\tS(t+1)=0$ and $\lambda_t\neq t+1$, then $\lambda_{t+1}=t+2$ from \eqref{def:lambda_t}. Next, if $\tS(t+1)=0$ and $\lambda_t= t+1$, then $\lambda_{t+1}=t+1=\lambda_t$ from \eqref{def:lambda_t}. Finally, if $\tS(t+1)>0$, then by Lemma \ref{lem:main}, we have $\lambda_{t+1}=\lambda_{t}$, and \eqref{lambdaupdate} follows. The proof is now complete.
	
	\section{Proofs of Results in Section~\ref{sec:asymptotics}}\label{sec:AppThm3}
	
	In this appendix, we let $M=N+H$ and $Y^j_k$ for $k=1,2,\ldots$ to be i.i.d.\ random variables such that 
	$$Y^j_k=\begin{cases}
	1 \quad \text{if $X_k\in I^N_j$,}\\
	0 \quad \text{if $X_k\not\in I^N_j$}.\\
	\end{cases}$$
	Recalling \eqref{muMLE}, we then have
	$$g_N(x)=\E[Y^j_1]$$ 
	and the regularized sample mean $\widehat{g}_N^{1:i}(x)$ defined in \eqref{eqn:regularized} can be written as
	$$\widehat{g}_N^{1:i}(x)=\frac{\sum_{k=1}^{i} Y^j_k+R}{i+MR}$$
	where $j$ is the unique integer such that $x\in I^N_j$. We first study the error bounds for the regularized sample mean $\widehat{g}_N^{1:i}(x)$. We then derive error bounds for the instantaneous sample log-likelihood ratio $\log\frac{\widehat{g}_N^{1:i}(x)}{{f}_N(x)}$. Finally we combine all the results together to derive an error bound for the growth rate of the BG-CuSum statistic.
	\subsection{Proof of Proposition~\ref{prop:ARL}}
	Following arguments identical to those in Chapter 2 of \cite{siegmund85}, it can be shown that
	\begin{align}\label{ARL1}
		\ARL(\widetilde{\tau}( {b}))&=\frac{\E{\infty}[\zeta( {b})]}{\P{\infty}(\tS(\zeta( {b}))\geq {b})}.
	\end{align}
	In order to obtain a lower bound for the ARL, we first note that for any $b>0$, we have
	\begin{align}\label{ARL_n}
		\E{\infty}[\zeta( {b})] \geq \E{\infty}[\zeta(0)]=1.
	\end{align}		
	We also have
	\begin{align}
		\P{\infty}(\tS(\zeta( {b}))\geq  {b})
		&=\P{\infty}(\sum_{i=1}^{\zeta( {b})}\log\frac{\widehat{g}_N^{1:i-1}(X_i)}{f_N(X_i)}\geq b)\label{explain_change_point}\\
		&=\P{\infty}(\prod_{i=1}^{\zeta( {b})}\frac{\widehat{g}_N^{1:i-1}(X_i)}{f_N(X_i)}\geq e^{b})\nonumber\\
		&=\sum_{t=1}^\infty\P{\infty}(E_t),\label{PinftyE_t}
	\end{align}
	where $E_t$ is the event $\left\{\prod_{i=1}^{t}\frac{\widehat{g}_N^{1:i-1}(X_i)}{f_N(X_i)}\geq e^ {b},\zeta( {b})=t\right\}$. The equality in \eqref{explain_change_point} follows from Theorem~\ref{thm:Supdate} and noting that since $\zeta(b)$ is the first time $\tS(t)$ exceeds $b$ or falls below $0$, $\lambda_i= \lambda_0=1$ for all $i=1,\ldots,\zeta(b)$ under the event $\tS(\zeta(b))\geq b$. 
	
	Let $J_i\in \{1,\ldots,N+H\}$ denote the index of the bin in $\{I^N_j\}_{j=1}^{N+H}$ that $X_i$ falls into. Let $F_t$ be the event such that $\{J_i\}_{i=1}^{t}\in F_t$ if and only if $\{X_i\}_{i=1}^{t}\in E_t$. For any $i\geq 1$ and any sequence $(x_1, \ldots, x_i)$ with corresponding bin indices $(j_1, \ldots, j_i)$, let $q(j_i\mid j_{1:i-1}) = \widehat{g}_N^{1:i-1}(x_i)$. From the Kolmogorov Extension Theorem\cite{bhattacharya2007basic}, there exists a probability measure $\bbQ$ on $\{1,\ldots,N+H\}^\mathbb{N}$ such that $\bbQ(J_1=j_1,\ldots,J_t=j_t) = \prod_{i=1}^t q(j_i \mid j_{1:i-1})$ for all $t \geq 1$. We then have
	\begin{align*}
		\P{\infty}(E_t)&=\E{\infty}[\mathds{1}_{E_t}]\\
		&\leq\E{\infty}[e^{-b}\prod_{i=1}^{t}\frac{\widehat{g}_N^{1:i-1}(X_i)}{f_N(X_i)}\mathds{1}_{E_t}]\\
		&\leq e^{-b}\sum_{\{j_i\}\in F_t} \prod_{i=1}^t q(j_i \mid j_{1:i-1})\\
		&=e^{-b}\bbQ(F_t).
	\end{align*}
	From \eqref{PinftyE_t}, we obtain 
	\begin{align}
		\P{\infty}(\tS(\zeta(b))\geq b)
		&\leq e^{- {b}}\sum_{t=1}^\infty\bbQ(F_t)\nonumber\\
		&\leq e^{- {b}}.\label{ARL_d}
	\end{align}
	where the final inequality follows because $\{F_t\}_{t\in\mathbb{N}}$ are mutually exclusive events. Thus, from \eqref{ARL1}, \eqref{ARL_n} and \eqref{ARL_d}, we have 
	\begin{align}
		\ARL(\widetilde{\tau}( {b}))&=\frac{\E{\infty}[\zeta( {b})]}{\P{\infty}(\tS(\zeta( {b}))\geq {b})}\geq e^{ {b} }. 
	\end{align}
	The proof is now complete.
	\subsection{Proof of Proposition~\ref{prop:errorbound_for_mu}}
	In the design of the BG-CuSum test statistic, we replace the maximum likelihood estimator $g_N^{1:i}$ with its regularized sample mean $\widehat{g}_N^{1:i}$. In this subsection, we derive a bound for the probability that $\widehat{g}_N^{1:i}(x)$ deviates from $g_N(x)$ by at least $\epsilon > 0$. We start with a few elementary lemmas. 
	
	\begin{Lemma_A}\label{lem:estimate_near_mean}
		For any $\epsilon \geq 0$, $N\geq 1$, and $x\in\mathbb{R}$, we have
		\begin{align*}
			\P{1}(\left|\widehat{g}_N^{1:i}(x)-\frac{\E[\sum_{k=1}^i Y^j_k]+R}{i+MR}\right|\geq \epsilon)    &\leq2e^{-2i\epsilon^2}
		\end{align*}
		where $j$ is the unique integer such that $x\in I^N_j$.	
	\end{Lemma_A}
	\begin{IEEEproof}
		By applying Hoeffing's inequality\cite{hoeffding1963probability}, we obtain
		\begin{align*}
			&\P{1}(\left|\widehat{g}_N^{1:i}(x)-\frac{\E[\sum_{k=1}^i Y^j_k]+R}{i+MR}\right|\geq \epsilon)  \\ 
			&=\P{1}(\left|\frac{\sum_{k=1}^i Y^j_k+R}{i+MR}-\frac{\E[\sum_{k=1}^i Y^j_k]+R}{i+MR}\right|\geq \epsilon)\\
			&=\P{1}(\left|\frac{\sum_{k=1}^i Y^j_k+R}{i}-\frac{\E[\sum_{k=1}^i Y^j_k]+R}{i}\right|\geq \frac{i+MR}{i}\epsilon)\\
			&\leq 2e^{-2i\left(\frac{i+MR}{i}\right)^2\epsilon^2}\leq 2e^{-2i\epsilon^2},
		\end{align*}
		and the proof is complete.
	\end{IEEEproof}
	
	\begin{Lemma_A}\label{lem:mean_near_true}
		For any $\epsilon \in (0,1)$ and $x\in\mathbb{R}$, there exists $i_0\in\mathbb{N}$ such that for all $i\geq i_0$, we have
		\begin{align*}
			\left|\frac{\E[\sum_{k=1}^i Y^j_k]+R}{i+MR}-g_N(x)\right|< \epsilon
		\end{align*}
		where $j$ is the unique integer such that $x\in I^N_j$.
	\end{Lemma_A}
	\begin{IEEEproof}
		For $i\geq(1-\epsilon)MR/\epsilon$, we have \\
		%<*tag:r1c51>
		\begin{align*}
			\left|\frac{\E[\sum_{k=1}^i Y^j_k]+R}{i+MR}-g_N(x)\right|&=\left|\frac{i g_N(x)+R}{i+MR}-g_N(x)\right|\\
			&=\left|\frac{R-MR g_N(x)}{i+MR}\right|\\
			&=\frac{R}{i+MR}\left|1-M g_N(x)\right|\\
			&\leq\frac{MR}{i+MR}\\
			&\leq\epsilon,
		\end{align*}
		%</tag:r1c51>
		and the proof is complete.
	\end{IEEEproof}
	
	Putting everything together, we now proceed to the proof of Proposition~\ref{prop:errorbound_for_mu}. Given $0<\epsilon<1$, by Lemma \ref{lem:mean_near_true} there exists $i_1>0$ such that for all $i\geq i_1$, we have %<*tag:r1c52>
	\begin{align*}
		\left|\frac{\E[\sum_{k=1}^i Y^j_k]+R}{i+MR}-g_N(x)\right|< \frac{\epsilon}{2}.
	\end{align*}%</tag:r1c52>
	Therefore, we obtain
	%<*tag:r1c53>
	\begin{align*}
		& \P{1}(\left|\widehat{g}_N^{1:i}(x)-g_N(x)\right|\geq \epsilon)\\
		&=  \P{1}(\left|\widehat{g}_N^{1:i}(x)-\frac{\E[\sum_{k=1}^i Y^j_k]+R}{i+MR}+\frac{\E[\sum_{k=1}^i Y^j_k]+R}{i+MR}-g_N(x)\right|\geq \epsilon)\\
		&\leq  \P{1}(\left|\widehat{g}_N^{1:t}(x)-\frac{\E[\sum_{k=1}^i Y^j_k]+R}{i+MR}\right|+\left|\frac{\E[\sum_{k=1}^i Y^j_k]+R}{i+MR}-g_N(x)\right|\geq \epsilon)\\
		&\leq  \P{1}(\left|\widehat{g}_N^{1:i}(x)-\frac{\E[\sum_{k=1}^t Y^j_k]+R}{i+MR}\right|\geq \frac{\epsilon}{2})\\
		&\leq  2e^{-\frac{1}{2}i\epsilon^2},
	\end{align*}
	%</tag:r1c53>
	where the last inequality follows from Lemma \ref{lem:estimate_near_mean}, and the proposition is proved.

	\subsection{Proof of Proposition~\ref{prop:error_instant_statistic}}
	
	In this subsection, we use previous results on the regularized sample mean $\widehat{g}_N^{1:i}$ to study the instantaneous log-likelihood ratio $\log\frac{\widehat{g}_N^{1:i}(x)}{{f}_N(x)}$ used in the BG-CuSum test statistic.
	
	\begin{Lemma_A}\label{lem:error_bound_log}
		For any $\epsilon \in (0,1)$, there exists a $t_2\in\mathbb{N}$ such that for all $i\geq t_2$ and any $x\in\mathbb{R}$, we have
		\begin{align*}
			\P{1}(\left|\log\widehat{g}_N^{1:i}(x)-\log g_N(x)\right|\geq \epsilon) \leq2e^{-i(g_N^{\min}\epsilon)^2/8}
		\end{align*}
		where $g_N^{\min}=\min_x g_N(x)$.
	\end{Lemma_A}
	\begin{IEEEproof}
		
		Using the inequality $\log x \leq x-1$, we obtain for each $x\in\mathbb{R}$,
		\begin{align*}
			\P{1}(\log \frac{\widehat{g}_N^{1:i}(x)}{g_N(x)} \geq \epsilon)
			&\leq \P(\widehat{g}_N^{1:i}(x) - g_N(x) \geq \epsilon g_N(x))
		\end{align*}
		and
		\begin{align*}
			\P{1}(\log \frac{g_N(x)}{\widehat{g}_N^{1:i}(x)} \geq \epsilon)
			&\leq \P(g_N(x) -\widehat{g}_N^{1:i}(x) \geq  \frac{\epsilon}{1+\epsilon}g_N(x) ).
		\end{align*}
		Therefore, we have
		
		\begin{align*}
			&\P{1}(|\log \widehat{g}_N^{1:i}(x)-\log g_N(x) |\geq \epsilon)\\
			&=\P{1}(\log \frac{\widehat{g}_N^{1:i}(x)}{ g_N(x)} \geq \epsilon)+\P(\log \frac{ g_N(x)}{\widehat{g}_N^{1:i}(x)} \geq \epsilon)\\
			&\leq \P{1}(\widehat{g}_N^{1:i}(x) -  g_N(x) \geq \epsilon  g_N(x))\\
			&\quad+\P{1}( g_N(x) -\widehat{g}_N^{1:i}(x) \geq  \frac{\epsilon}{1+\epsilon} g_N(x))\\
			&\leq \P{1}\left(|\widehat{g}_N^{1:i}(x) -  g_N(x)|\geq \min\left(\epsilon  g_N(x),\frac{\epsilon}{1+\epsilon} g_N(x)\right)\right)\\
			&\leq \P{1}\left(|\widehat{g}_N^{1:i}(x) -  g_N(x)|\geq \frac{1}{2}\epsilon g_N(x)\right)\\
			&\leq 2e^{-i( g_N(x)\epsilon)^2/8}\leq 2e^{-i( g_N^{\min}\epsilon)^2/8},
		\end{align*}
		
		where the last inequality follows from Proposition~\ref{prop:errorbound_for_mu} for all $i \geq t_1^x$, where $t_1^j\in\Nat$ is chosen to be sufficiently large. Taking $t_2=\max_{x\in\mathbb{R}} t_1^x$, the lemma follows.	
	\end{IEEEproof}
	We now proceed to the proof of Proposition~\ref{prop:error_instant_statistic}. Using the law of total probability and Lemma~\ref{lem:error_bound_log}, we have for $i \geq t_2$, where $t_2$ is as given in Lemma~\ref{lem:error_bound_log},
	\begin{align*}
		&\P{1}(\left|\log\frac{\widehat{g}_N^{1:i}(X_{i+1})}{f_N(X_{i+1})}-\log\frac{ g_N(X_{i+1})}{f_N(X_{i+1})}\right|\geq \epsilon)\\
		&=\sum_{j=1}^{M}\P{1}(\left|\log\widehat{g}_N^{1:i}(X_{i+1})-\log g_N(X_{i+1})\right|\geq \epsilon){X_{i+1}\in I^N_j}\\
		&\quad\times\P{1}(X_{i+1}\in I^N_j)\\
		&=\sum_{j=1}^{M}\P{1}(\left|\log\widehat{g}_N^{1:i}(X_{i+1})-\log g_N(X_{i+1})\right|\geq \epsilon){X_{i+1}\in I^N_j}\\
		&	\quad\times\P{1}(X_{i+1}\in I^N_j)\\
		&=\sum_{j=1}^{M}\P{1}(\left|\log\widehat{g}_N^{1:i}(X_{i+1})-\log g_N(X_{i+1})\right|\geq \epsilon)\\
		&\quad\times\P{1}(X_{i+1}\in I^N_j)\\
		&\leq \sum_{j=1}^{M} 2e^{-\frac{i}{8}( g_N^{\min}\epsilon)^2} \P{1}(X_{i+1}\in I^N_j)\\
		&=  2e^{-i( g_N^{\min}\epsilon)^2/8} ,
	\end{align*}
	and the proof is complete.

	\subsection{Proof of Proposition~\ref{thm:rate_of_growth}}
	In this subsection, we use results derived in the previous two subsections to study the growth rate $\tS(t)/t$ of the BG-CuSum test statistic.
	
	\begin{Lemma_A}\label{lem:error_true_to_statistic}
		For any $\epsilon \in (0,1)$, there exists a $t_3\in\mathbb{N}$ such that for all $t\geq t_3$,  we have
		\begin{align*}
			\P{1}(\left|\frac{1}{t}\sum_{i=1}^t\log\frac{\widehat{g}_N^{1:i-1}(X_{i})}{ g_N(X_i)}\right|\geq \epsilon) \leq c_2 e^{-c_1 t^{1-\epsilon}},
		\end{align*}
		where $c_1, c_2$ are positive constants.
	\end{Lemma_A}
	
	\begin{IEEEproof}
		Let $l = \ceil{t^{1-\epsilon}}$. For all $i \leq l$, we have from \eqref{eqn:regularized},
		\begin{align*}
			\wmu{1:i-1}(X_i) \geq \frac{R}{l + MR} \quad \text{$\P_1$-a.s.},
		\end{align*}
		which yields
		\begin{align}
			\left| \log\wmu{1:i-1}(X_i) - \log  g_N(X_i) \right| \leq \log\frac{l + MR}{R} + |\log g^{\min}_N|,  
		\end{align}
		where $g^{\min}_N = \min_j  g_N(j) > 0$ since $f_N(j) = 1/N$ for all $j$ and is absolutely continuous w.r.t.\ $ g_N$. There exists $t_3 \in\Nat$ such that for all $t \geq t_3$, 
		\begin{align*}
			&\frac{l}{t} \left(\log\frac{l + MR}{R} + |\log g^{\min}_N|\right)\\ 
			&\leq \frac{t^{1-\epsilon}+1}{t} \left(\log\frac{t^{1-\epsilon}+1+ MR}{R} + |\log g^{\min}_N|\right) \\
			&\leq \left(t^{-\epsilon}+\frac{1}{t} \right)\left(\log\frac{t^{1-\epsilon}+1+ MR}{R} + |\log g^{\min}_N|\right) \leq \frac{\epsilon}{2}.
		\end{align*}
		We then obtain for $t \geq t_3$,
		\begin{align*}
			&\P{1}(\left|\frac{1}{t}\sum_{i=1}^t\log\frac{\widehat{g}_N^{1:i-1}(X_{i})}{ g_N(X_i)}\right|\geq \epsilon) \\
			&\leq \P{1}(\left|\frac{1}{t-l}\sum_{i=l+1}^t\log\frac{\widehat{g}_N^{1:i-1}(X_{i})}{ g_N(X_i)}\right| + \frac{l}{t} \left(\log\frac{l + MR}{R} + |\log g^{\min}_N|\right) \geq \epsilon) \\
			& \leq \P{1}(\left|\frac{1}{t-l}\sum_{i=l+1}^t\log\frac{\widehat{g}_N^{1:i-1}(X_{i})}{ g_N(X_i)}\right|  \geq \frac{\epsilon}{2}) \\
			& \leq \sum_{i=l+1}^t \P{1}(\left|\log\widehat{g}_N^{1:i-1}(X_{i})-\log g_N(X_i)\right|  \geq \frac{\epsilon}{2}) \\
			& \leq 2\sum_{i=l+1}^t e^{-i( g_N^{\min}\epsilon)^2/8} \\
			& \leq c_2 e^{- c_1 l} \leq c_2 e^{-c_1 t^{1-\epsilon}},
		\end{align*}
		where the penultimate inequality follows from Proposition~\ref{prop:error_instant_statistic}, $c_1 = ( g_N^{\min}\epsilon)^2/8$, and $c_2 = 2(1 - e^{-c_1})^{-1}$. The lemma is now proved. 
	\end{IEEEproof}
	Finally, we are ready to prove Proposition~\ref{thm:rate_of_growth}. We begin by showing that $\frac{1}{t}\sum_{i=1}^t\log\frac{\widehat{g}_N^{1:i-1}(X_{i})}{ g_N(X_i)}$ converges to zero r-quickly under the distribution $\P_{1}$. For any $\epsilon \in (0,1)$, let 
	\begin{align*}
		L_\epsilon=\sup\left\{t:\left|\frac{1}{t}\sum_{i=1}^t\log\frac{\widehat{g}_N^{1:i-1}(X_{i})}{ g_N(X_i)}\right|> \epsilon\right\}.
	\end{align*}
	We have
	\begin{align*}
		\E{1}[L_\epsilon]&=\sum_{n=1}^\infty \P{1}(L_\epsilon\geq n)\\
		&=\sum_{n=1}^\infty \P{1}(\left|\frac{1}{t}\sum_{i=1}^t\log\frac{\widehat{g}_N^{1:i-1}(X_{i})}{ g_N(X_i)}\right|\geq \epsilon, \text{for some $t\geq n $})\\
		&\leq\sum_{n=1}^\infty\sum_{t=n}^\infty \P{1}(\left|\frac{1}{t}\sum_{i=1}^t\log\frac{\widehat{g}_N^{1:i-1}(X_{i})}{ g_N(X_i)}\right|\geq \epsilon)\\
		&=\sum_{n=1}^\infty\sum_{t=n}^\infty c_2 e^{-c_1 t^{1-\epsilon}}\\
		&=c_2\sum_{n=1}^\infty n e^{-c_1 n^{1-\epsilon}}<\infty,
	\end{align*}
	where the penultimate equality follows from Lemma~\ref{lem:error_true_to_statistic}, and $c_1,c_2$ are positive constants. Thus, $\frac{1}{t}\sum_{i=1}^t\log\frac{\widehat{g}_N^{1:i-1}(X_{i})}{ g_N(X_i)}$ converges to zero $r$-quickly for $r=1$ under the distribution $\P_{1}$. 
	
	Since $\E{1}[\left(\log \frac{g_N(X)}{f_N(X)}\right)^2]<\infty$, from Theorem~2.4.4 of \cite{tartakovsky2014sequential}, $\frac{1}{t}\sum_{i=1}^t\log\frac{g_N(X_i)}{f_N(X_i)}$ converges to $\KLD{g_N}{f_N}$ $r$-quickly for $r=1$ under the distribution $\P_{1}$. Therefore, $\frac{1}{t}\sum_{i=1}^t\log\frac{\widehat{g}_N^{1:i-1}(X_{i})}{ f_N(X_i)}$ converges to  $\KLD{g_N}{f_N}$ $r$-quickly for $r=1$ under $\P_{1}$ and the proof is complete. 
	
	\subsection{Proof of Proposition~\ref{thm:add_arl_shat}}
	Let $$\widehat{T}=\inf\{t:\sum_{i=1}^t\log\frac{ \widehat{g}^{1:i-1}_N(X_i)}{f_N(X_i)}>b\}.$$ Using Proposition~\ref{thm:rate_of_growth} and Corollary~3.4.1 in~\cite{tartakovsky2014sequential}, we obtain $$\E{1}[\widehat{T}]\sim\frac{b}{\KLD{ g_N}{f_N}} \quad\text{as $b\to\infty$.}$$  Using arguments similar to those that led to Eq~\eqref{ARL_d}, we obtain $$\P_{\infty}(\widehat{T}<\infty)=\sum_{t=1}^\infty\P_{\infty}(\widehat{T}=t)\leq e^{-b}.$$
	
	Applying results from Theorem 6.16 in~\cite{poor2009quickest} to translate our understanding of $\widehat{T}$ onto $\widehat{\tau}(b)$, we obtain $$\ARL(\widehat{\tau}(b)) \geq \frac{1}{\P_{\infty}(\widehat{T}<\infty)}= e^b$$ and that $$\text{WADD}(\widehat{\tau}(b))\leq \E{1}[\widehat{T}]\sim\frac{b}{\KLD{ g_N}{f_N}}\quad\text{as $b\to \infty$.}$$
	
	For the case where $f$ and $g$ are discrete distributions, we have $f_N =f$ and $g_N=g$, so that these bounds coincide with the bounds for the CuSum stopping time when both $f$ and $g$ are known. Thus, for this case, the test is asymptotically optimal and the theorem is now proved.
	
	\bibliographystyle{IEEEtran}
	\bibliography{IEEEabrv,StringDefinitions,refs}

%% file: supplementary_materials_arxiv.tex
	\maketitle
	In this supplementary material, we give an exact characterization of $N$ such that a distribution $g$ is distinguishable from $f$ w.r.t. $N$. We also provide an example to illustrate how $N$ can be determined if additional moment information is available.
	\section{Properties of \texorpdfstring{$D(f,N)$}{D}}\label{sec:distinguishable_distributions}
	
	In the design of the BG-CuSum test, we assume that the post-change distribution $g$ is distinguishable from ${f}$ w.r.t.\ $N$. In this section, we derive some properties of $D(f,N)$ and give an example on how to choose $N$ for a particular family of post-change distributions.
	
	Let $F_c$ be the cumulative density function (cdf) of the continuous part of the pre-change distribution ${f}$. Let $G_c$ be the cdf of the continuous part of the unknown post-change distribution $g$. Let $I$ be the image under $F_c$ of the zero set of $F_c-G_c$ defined as: 
	\begin{align}\label{eqn:I}
	I=\left\{F_c(x) : \ F_c(x)-G_c(x)=0,x\in\mathbb{R}\cup\{-\infty,\infty\}\right\},
	\end{align}
	where the terms $F_c(\infty),F_c(-\infty),G_c(\infty),G_c(-\infty)$ are defined to be the limits of $F_c$ and $G_c$ as $x$ tends to $\infty$ or $-\infty$ respectively. 
	
	We first begin by deriving a necessary and sufficient condition for $g$ to be distinguishable from ${f}$ w.r.t. $N$.
	
	\begin{Proposition_A}\label{prop:exact}
		The distribution $g\in D({f},N)$ if and only if $p_h\neq q_h$ for some $h\in\{0,...,H\}$ or
		\begin{align}
		\left\{\tfrac{i}{N}  : 0\leq i \leq N \right\}\setminus I \neq \emptyset.\label{iNI}
		\end{align}
	\end{Proposition_A}
	
	\begin{IEEEproof}
		Let the intervals $(a_0,a_1],(a_1,a_2],...,(a_{N-1},a_N)$ with $a_0=-\infty,a_N=\infty$ be such that for each $i\in\{1,...,N-1\}$, $F_c(a_i)=\tfrac{i}{N}.$
		
		If $p_h\neq q_h$ for some $h\in\{0,...,H\}$, then $g\in D({f},N)$ trivially. Suppose now that $p_h=q_h$ for all $h\in\{0,...,H\}$, and \eqref{iNI} holds. There exists $a_i$ such that $F_c(a_i)=i/N$ and $G_c(a_i)\neq i/N$. This implies that $\sum_{j=1}^i{f}_N(a_j)\neq \sum_{j=1}^ig_N(a_j)$. Therefore, there exists at least one $j$ such that ${f}_N(a_j)\neq g_N(a_j)$. Hence $g$ is distinguishable from ${f}$ w.r.t.\ $N$. 
		
		On the other hand, suppose now that $g\in D({f},N)$ and $p_h=q_h$ for all $h\in\{0,...,H\}$. Then for some $j\in\{1,...,N-1\}$,
		\begin{align}
		F_c(a_j)-F_c(a_{j-1})\neq G_c(a_j)-G_c(a_{j-1}).\label{eqn:unequal}
		\end{align}  
		Letting $j_0$ be the smallest of such $j$ satisfying \eqref{eqn:unequal}, we have $F_c(a_{j_0})\neq G_c(a_{j_0})$. Therefore, $a_{j_0}\notin I$. Since $F_c$ is injective, $j_0/N\notin I.$ Therefore the left hand side of \eqref{iNI} is non-empty. The proof is now complete.
	\end{IEEEproof}
	
	An easy application of Proposition~\ref{prop:exact} relates the number of elements of the set $I$ and $N$ for $g$ to be distinguishable from ${f}$ w.r.t. $N$.
	
	\begin{Corollary_A}\label{cor:upperbound}
		If $|I|$ is finite, then $g\in D({f},N)$ for all $N\geq |I|$.
	\end{Corollary_A}
	\begin{IEEEproof}
		%Since $0,1\in I$, there are $N-2$ point in $T\setminus\{0,1\}$. Let $(-\infty,a_1),[a_1,a_2),...,[a_{N-1},\infty)$ be $N$ intervals such that $F(a_i)=\tfrac{i}{N}$ for $i=1,...,N-1$ and $a_0=-\infty$. Using the pigeon-hole principle, there exists $i_0\in\{1,...,N-1\}$ such that $i_0=\argmin \{j:a_{j}\notin T\setminus\{0,1\}\}$. 
		%Therefore, we have $F(a_{i_0})\neq G(a_{i_0})$. Furthermore, we have $G(a_{i_0})-G(a_{i_0-1})\neq F(a_{i_0})-F(a_{i_0-1})=\tfrac{1}{N}$. Thus, ${f}$ is distinguishable from $g$ with respect to $N$.
		For any $N\geq |I|$, counting the number of elements in each set we obtain 	
		$$\left|\left\{\tfrac{i}{N}  : 0\leq i \leq N \right\}\setminus I\right|\geq 1.$$
		From Proposition \ref{prop:exact}, we have $g\in D({f},N)$, and the corollary follows.
	\end{IEEEproof}
	
	As an example, if the post-change distribution $g$ is $f$ shifted in mean, then $|I|=1$ and we can choose $N=1$. In general, $N$ can be chosen based on prior statistical information about $g$. As an illustration, we consider post-change distributions that satisfy the following assumption in the remainder of this section.
	\begin{Assumption}\label{assumpt:moments}
		For some positive integer $k$ and $\epsilon>0$, we have $|\E^{g}[X^{k}]-\E^{{f}}[X^{k}]|>\epsilon$. Furthermore, there exist $C, \xi >0$ such that $f_c(x)\leq C |x|^{-k-1-\xi}$ and $g_c(x)\leq C |x|^{-k-1-\xi}$ for all $x\in\Real$. 
	\end{Assumption}
	Note that Assumption~\ref{assumpt:moments} does not require us to know the $k$-th moment of $g$. In the following, we present a result that allows us to derive an algorithm for selecting $N$ so that $g\in D({f},N)$. It gives a lower bound on $|I|$ if the first $k-1$ moments of ${f}$ and $g$ are equal. 
	
	\begin{Proposition_A}\label{prop:moments_lower_bound}
		Suppose $g_c\ne {f_c}$, the first $k-1$ moments of ${f_c}$ and $g_c$ are equal. Then, $|I|\geq k$.
	\end{Proposition_A}
	\begin{IEEEproof}
		The claim is trivial if $k < 2$. Therefore, we consider only the case where $k\geq 2$, and proceed by contradiction. Suppose $|I|<k $. 
		
		We define a sign change to be a compact interval $[a,b]$ (with possibly $b=a$, in which case $[a,b]$ is a singleton set) such that $F_c(x)-G_c(x)=0$ for all $x\in[a,b]$, and there exist $x_1 < a \leq b < x_2$ such that $(F_c(x_1)-G_c(x_1))(F_c(x_2)-G_c(x_2)) < 0$. Then, from the definition of $I$ in \eqref{eqn:I}, since $F_c(x)-G_c(x)$ has at most $|I|-2<k-2$ sign changes,  there exists a degree $k-1$ polynomial $P(x)=\sum_{i=0}^{k-1}a_i x^i$ such that its derivative $P'(x)$ satisfies $P'(x)\left(F_c(x)-G_c(x)\right)>0$ for all $x\in Z=\mathbb{R}\setminus \{x\in\mathbb{R} :  F_c(x)=G_c(x)\}$. Note that since $g_c\ne{f_c}$ according to our definition, the set $Z$ has non-zero Lebesgue measure. Integrating by parts, we have 
		\begin{align*}
		0&<\int_{-\infty}^{\infty} P'(x)\left(F_c(x)-G_c(x)\right)\ \ud x \\
		&=-\int_{-\infty}^{\infty} P(x)(f_c(x)-g_c(x))\ \ud x \\
		&=-\sum_{i=0}^{k-1} a_i \int_{-\infty}^{\infty} x^i(f_c(x)-g_c(x))\ \ud x=0,
		\end{align*}
		where the last equality follows from the assumption that the first $k-1$ moments of ${f_c}$ and $g_c$ are equal. This gives us a contradiction. Therefore, $|I|\geq k$, and the proof is complete.
	\end{IEEEproof}
	
	The following theorem gives us a method to search for $N$ such that $g\in D({f},N)$. 
	
	\begin{Theorem_A}\label{thm:moments}
		Suppose Assumption~\ref{assumpt:moments} holds. If $g\notin D({f},N)$, we have $ m_N\leq\E^{g}[X^{k}]\leq M_N$, where  
		\begin{align}
		M_N &=p_0\left(\ofrac{N}\sum_{i=2}^{N-1}\max_{x\in I^N_i}(x^{k}) + \int_{I^N_1\cup I^N_N}\max\left(\tfrac{Cx^k}{|x|^{k+1+\xi}},0\right) \ud x\right)+\sum_{h=1}^H p_h\theta_h^k, \label{M_N}\\
		m_N &=p_0\left(\ofrac{N}\sum_{i=2}^{N-1}\min_{x\in I^N_i}(x^{k}).\right)+\sum_{h=1}^H p_h \theta_h^k \label{m_N}
		\end{align}	
		Furthermore, $\displaystyle\lim_{d\to\infty}M_{2^d}=\lim_{d\to\infty}m_{2^d}=\E^{{f}}[X^{k}].$
	\end{Theorem_A}
	\begin{IEEEproof}
		See Section~\ref{sec:appthm2}.
	\end{IEEEproof}
	Theorem~\ref{thm:moments} shows that there exists a sufficiently large $N$ so that $g\in D({f},N)$. To determine a suitable $N$, we note from Corollary \ref{cor:upperbound} that $g$ is  distinguishable from ${f}$ w.r.t.\ $N$ for any $N\geq |I|$. From Proposition \ref{prop:moments_lower_bound}, a candidate to start the search of $N$ such that $g\in D({f},N)$ would be $N=k$. A procedure to find $N$ so that $g\in D({f},N)$ is given in Algorithm~\ref{alg:ComputeN}, which is guaranteed to stop after a finite number of iterations due to Theorem~\ref{thm:moments}.
		\begin{algorithm}[!htb]
		\caption{Given Assumption~\ref{assumpt:moments}, compute \texorpdfstring{$N$}{N} so that \texorpdfstring{$g\in D({f},N)$}{distinguishability}.}\label{alg:ComputeN}
		\begin{algorithmic}[1]
			\State $\textbf{Initialize: }$
			\State Set $N:=k$
			\While {$\left(M_N>\E^{{f}}[X^{k}]+\epsilon\text{ or } m_N<\E^{{f}}[X^{k}]-\epsilon\right)$}
			\State $N:= N+1$
			\State Compute the bins $I_i^N$, $i=1,\ldots,N+H$, according to Definition~\ref{def:bins}.
			\State Compute $M_N$ and $m_N$ according to \eqref{M_N} and \eqref{m_N}, respectively.
			\EndWhile\\
			\Return $N$
		\end{algorithmic}
	\end{algorithm}
\section{Proof of Theorem \ref{thm:moments}}\label{sec:appthm2}
%\red{[I didn't go through the details. Please go over and update notations and references.]}		
We break the proof into two parts. In Proposition \ref{prop:moments_distinguishable}, we derive a upper bound $M_N$ and lower bound $m_N$ on the $k$-th moment of $g$ if $g\notin D({f},N)$. Then, in Proposition~\ref{prop:converge}, we show that the sub-sequence $M_{2^u}$ and $m_{2^u}$ converges to $\E^{{f}}[X^{k}]$ as $u\to\infty$. We let $\mathds{1}_A(x)$ be the indicator function for the set $A$.
		\begin{Proposition_A}\label{prop:moments_distinguishable}
			Under the setup of Theorem~\ref{thm:moments}, we have $ m_N\leq\E^{g}[X^{k}]\leq M_N$.
		\end{Proposition_A}
		
		\begin{IEEEproof}
			Since $g\notin D({f},N)$, $p_h=q_h$ for $h\in\{0,...,H\}$ and we have for $j=1,...,N$,
			\begin{align*}
			\int_{I^N_j}g_c(x)\ud x=\int_{I^N_j}f_c(x)\ud x = \frac{1}{N},
			\end{align*}
			which yields
			\begin{align*}
			M_N
			&=p_0\left(\sum_{j=2}^{N-1}\max_{x\in I^N_j}(x^{k})\int_{I^N_j}g_c(x) \ud x + \int_{I^N_1\cup I^N_N}\max\left(\frac{Cx^{k}}{|x|^{k+1+\xi}},0\right) \ud x\right)+\sum_{h=1}^Hp_h\theta_h^k\\
			&\geq p_0\left(\sum_{j=1}^N\int_{I^N_j}x^{k}g_c(x) \ud x\right)+\sum_{h=1}^Hp_h\theta_h^k= \int_{\Real}x^{k}g(x) \ud x = \E^{g}[X^{k}].
			\end{align*}
			Similarly, we have
			\begin{align*}
			m_N
			&=p_0\left(\sum_{j=2}^{N-1}\min_{x\in I^N_j}(x^{k})\int_{I^N_j}g_c(x) \ud x\right)+\sum_{h=1}^Hp_h\theta_h^k\\
			&\leq\int_{\Real}x^{k}g(x) \ud x=\E^{g}[X^{k}],
			\end{align*}
			and the proof is complete.
		\end{IEEEproof}
		
		We require the use of the Dominated Convergence Theorem \cite{rudin1987real} to show that the upper bound $M_{2^u}$ converges to $\E^{g}[X^{k}]$. In the next lemma, we construct an integrable dominating function $v(x)$.
		\begin{Lemma_A}\label{lem:Integrable_v}
			Suppose Assumption~\ref{assumpt:moments} holds. For $i\geq 1$, let the intervals $J_i=(a_{i-1},a_{i}]$, $J_{-i}=(a_{-i},a_{-i+1}]$ and $J_0=\emptyset$ be defined such that
			\begin{align*}
			\int_{-\infty}^{a_0} f_c(x)\ \ud x=\frac{1}{2}\quad \text{and}\quad \int_{J_{i}}f_c(x)\ \ud x=\int_{J_{-i}}f_c(x)\ \ud x=\frac{1}{2^{i+1}}.
			\end{align*}
			Let $v(x)=\sum_{i=-\infty}^{\infty}\max_{x\in J_i}\{|x|^{k}\}\mathds{1}_{ J_i}(x) f_c(x)$, then $v(x)$ is integrable  
			$\int_{\mathbb{R}} v(x) \ud x <\infty$ 
		\end{Lemma_A}
		\begin{IEEEproof}
			It is either the case that $a_0<a_i\leq 0$ for all $i\in \mathbb{N}$ or there exists an $n$ such that  $a_i>0$ for all $i>n$. For the prior case, since $a_0 < a_i\leq 0$ we have for all $i\geq 1$, $0\leq \max_{x\in J_{i}}\{|x|^{k}\}\leq |a_0|^{k}$. Thus
			\begin{align*}
			\int_{a_0}^{\infty} v(x) \ud x= \sum_{i=1}^\infty \max_{x\in J_{i}}\{|x|^{k}\}\frac{1}{2^{i+1}}<\infty.
			\end{align*} 
			For the latter case, in order show that \begin{align*}\int_{a_0}^{\infty} v(x)\ud x<\infty,\end{align*} 
			we derive an upperbound for $\max_{x\in J_{i}}\{|x|^{k}\}$ so that $\sum_{i=n+1}^\infty \max_{x\in J_{i}}\{|x|^{k}\}\frac{1}{2^{i+1}}<\infty$. Using the assumption that $f_c(x)\leq C|x|^{-k-1-\xi}$ and $a_i>0$ for $i>n$, we obtain
			\begin{align}
			2^{-i-1}=\sum_{j=i}^\infty \frac{1}{2^{j+1}}&=\sum_{j=i}^\infty\int_{J_j}f_c(x)\ud x\nonumber \\
			&=\int_{a_i}^{\infty}f_c(x)\ud x \nonumber\\
			&\leq \int_{a_i}^\infty C|x|^{-k-1-\xi}\ dx\nonumber\\
			&=\frac{C}{k+\xi}{a_i}^{-k-\xi}. \label{eqn:aBound}
			\end{align}
			Using \eqref{eqn:aBound}, we obtain
			\begin{align*}
			2^{-i-1}&\leq \frac{C}{k+\xi}{a_i}^{-k-\xi}.
			\end{align*}
			Noting that $a_i>0$ for $i\geq n+1$, we have an upper bound for $a_i$
			\begin{align*}
			{a_i}&\leq \left(\frac{C}{k+\xi}\right)^\frac{1}{k+\xi}\left(2^{i+1}\right)^{\frac{1}{k+\xi}}\quad \text{for $i\geq n+1$}.
			\end{align*}
			Thus, we have an upper bound for $\max_{x\in J_{i}}\{|x|^{k}\}$ for $i\geq n+1$,
			\begin{align}
			\max_{x\in J_{i}}\{|x|^{k}\}=a_{i+1}^{k}&\leq \left(2^{\frac{k}{k+\xi}} \right)^{i+1}\left(\frac{C}{k+\xi}\right)^{\frac{k}{k+\xi}}\quad \text{for $i\geq n+1$}.\label{eqn:maxBound}
			\end{align} 
			Using the bound in \eqref{eqn:maxBound}, we are able to bound the integral $\int_{a_{0}}^{\infty}  v(x) \ud x$ by 
			\begin{align}
			&\int_{a_{0}}^{a_{n+1}}  v(x) \ud x+\int_{a_{n+1}}^{\infty}v(x) \ud x\nonumber\\ &=\int_{a_{0}}^{a_{n+1}}v(x) \ud x+\sum_{i={n+1}}^\infty\max_{x\in J_{i}}\{|x|^{k}\}2^{-i-1}\nonumber\\
			&\leq\int_{a_{0}}^{a_{n+1}}v(x) \ud x+\sum_{i={n+1}}^\infty\left(2^{\frac{k}{k+\xi}} \right)^{i+1}\left(\frac{C}{k+\xi}\right)^{\frac{k}{k+\xi}}2^{-i-1}\nonumber\\
			&\leq\int_{a_{0}}^{a_{n+1}}v(x) \ud x+\sum_{i={n+1}}^\infty\left(2^{-\frac{\xi}{k+\xi}} \right)^{i+1}\left(\frac{C}{k+\xi}\right)^{\frac{k}{k+\xi}}.\label{eqn:iNt}
			\end{align}
			Since $v(x)$ is bounded on the closed interval $[a_0,a_{n+1}]$, $\int_{a_{0}}^{a_{n+1}}v(x) \ud x$ is finite. Furthermore, there exist a positive real number $B_3$ such that 
			\begin{align*}
			\left(\frac{C}{k+\xi}\right)^\frac{k}{k+\xi}<B_3\quad \text{for $k\geq n+1$}.
			\end{align*} 
			Therefore, we bound the summation in \eqref{eqn:iNt} by 
			\begin{align*}
			\sum_{i={n+1}}^\infty\left(2^{-\frac{\xi}{k+\xi}} \right)^{i+1}\left(\frac{C}{k+\xi}\right)^{\frac{k}{k+\xi}}<B_3\sum_{i={n+1}}^\infty\left(2^{-\frac{\xi}{k+\xi}} \right)^{i+1}.
			\end{align*}
			Thus, we conclude that $\int_{a_{0}}^{\infty}  v(x) \ud x<\infty$. A similar argument can be used to show that $\int_{-\infty}^{a_0}v(x) \ud x <\infty$. Therefore, we have $\int_{\mathbb{R}}v(x) \ud x<\infty$.
		\end{IEEEproof}
		\begin{Lemma_A}\label{lem:dominate_v}
			Suppose Assumption 2 holds. For any $j\in \{2,...,2^u-1\}$, there exists an integer $i$ such that $I_j^{2^u}\subseteq J_{i}$.  Furthermore, for any positive integer $u$, we have
			\begin{align}
			&\left|\sum_{j=2}^{2^u-1}\max_{x\in I^{2^u}_j}(x^{k}) f_c(x)\mathds{1}_{I_j}(x)\right|\leq v(x) \quad\text{for $x\in\mathbb{R}$}. \label{eqn:summands}
			\end{align}
		\end{Lemma_A}
		\begin{IEEEproof}
			We will show that for any $j\in \{2,...,2^u-1\}$, there exists an integer $i$ such that $I_j^{2^u}\subseteq J_{i}$  by induction on $u$.
			For $u=2$, we can check that $I^{2^2}_{2}\subseteq J_{-1}$,$I^{2^2}_{3}\subseteq J_{1}$. Suppose the statement is true for $u=u_0$. For $u=u_0+1$, the intervals $\{I^{2^{u_0+1}}_j|j=1,...,2^{u_0+1}\}$ is a refinement of $\{I^{2^{u_0}}_j|j=1,...,2^{u_0}\}$. Hence, for each $j\in \{3,...,2^{u_0+1}-2\}$, there exist an integer $i$ such that $I_j^{2^{u_0+1}}\subseteq J_{i}$. It remains for us to check that $I^{2^{u_0+1}}_2\subseteq J_{-u_0}$ and that $I^{2^{u_0+1}}_{2^{u_0+1}-1}\subseteq J_{u_0}$. This is the case because the end-points of the intervals $a_i$ corresponds to point $\int_{a_i}^\infty f_c(x) \ud x=\frac{1}{2^{i+1}}$ if $i>0$ and $\int_{-\infty}^{a_i}f_c(x)\ud x=\frac{1}{2^{-i+1}}$ if $i<0$. So $\{J_j\}$ partitions the real line into smaller and smaller intervals towards infinity. For a fixed $u$, $I_2^{2^u}\subseteq J_{-(u-1)}$ and $I_{2^u-1}^{2^u}\subseteq J_{u-1}$.
			By mathematical induction, for any $j\in \{2,...,2^u-1\}$ there exist $i$ such that $I_j^{2^u}\subseteq J_i$ for $u\geq 2$.
			For a fixed $i$, for any $j$ such that $I_j^{2^u}\subseteq J_i$, we have
			\begin{align*}
			\max_{x\in I^{2^u}_j}(|x|^{k})\mathds{1}_{I^{2^u}_j}(x)\leq\max_{x\in J_i}(|x|^{k})\mathds{1}_{J_i}(x) \quad \text{for any $x\in\mathbb{R}$}.
			\end{align*}
			For each of the summands in \eqref{eqn:summands}, we have the following bound
			\begin{align*}
			\left|\max_{x\in I^{2^u}_j}(x^{k})\mathds{1}_{ I^{2^u}_j}(x)f_c(x)\right|&\leq \max_{x\in I^{2^u}_j}(|x|^{k})\mathds{1}_{ I^{2^u}_j}(x)f_c(x)\\
			&\leq \max_{x\in J_i}(|x|^{k})\mathds{1}_{ J_i}(x)f_c(x) \quad \text{for any $x\in\mathbb{R}$}.
			\end{align*}
			Putting everything together, we obtain
			\begin{align*}
			\left|\sum_{j=2}^{2^u-1}\max_{x\in I^{2^u}_j}(x^{k})f_c(x)\mathds{1}_{ I^{2^u}_j}(x)\right|&=\sum_{j=2}^{2^u-1}\left|\max_{x\in I^{2^u}_j}(x^{k})f_c(x)\mathds{1}_{ I^{2^u}_j}(x)\right|\\
			&\leq v(x)\ \text{for $x\in\mathbb{R}$}
			\end{align*} 
		\end{IEEEproof}
		\begin{Proposition_A}\label{prop:converge}
			Under the setup of Theorem~\ref{thm:moments}, we have $\displaystyle\lim_{u\to\infty}M_{2^u}=\lim_{u\to\infty}m_{2^u}=\E^{{f}}[X^{k+1}].$
		\end{Proposition_A}
		\begin{IEEEproof}
			We define the upper and lower bound functions $u_N(x)$ and $l_N(x)$ as  
			\begin{align}
			u_N(x)&=\max\left(\frac{Cx^{k}}{|x|^{-k-1-\xi}},0\right)\mathds{1}_{I^N_{1}}(x)\nonumber \\&\quad+\sum_{i=2}^N \max_{x\in I^N_{i}}(x^{k})f_c(x)\mathds{1}_{I^N_{i}}(x) \nonumber \\
			&\quad\quad+\max\left(\frac{Cx^{k}}{|x|^{-k-1-\xi}},0\right)\mathds{1}_{I^N_{N}}(x),\\
			l_N(x)&=\sum_{i=2}^N \min_{x\in I^N_{i}}(x^{k})f_c(x)\mathds{1}_{I^N_{i}}(x),
			\end{align}
			Note that $p_0\int_{\mathbb{R}}l_N(x)\ud x+\sum_{h=1}^Hp_h\theta_h^k=m_N$ and $p_0\int_{\mathbb{R}}u_N(x)\ud x+\sum_{h=1}^Hp_h\theta_h^k=M_N$. Furthermore, $l_N$ and $u_N$ converges pointwise to $x^{k}f_c(x)$ as $N$ tends to infinity. 
			Since $x^{k}f(x)$ is integrable and by the Monotone Convergence Theorem \cite{rudin1987real} , we have 
			\begin{align}
			\lim_{u\to\infty}m_{2^u}=\lim_{u\to\infty}p_0\int_{\mathbb{R}}l_{2^u}(x) \ud x+\sum_{h=1}^Hp_h\theta_h^k=\int_{\mathbb{R}}x^{k} f(x) \ud x =\E^{{f}}[X^{k}].\label{eqn:mlimit}
			\end{align}
			From Lemma \ref{lem:Integrable_v} and \ref{lem:dominate_v},  $\sum_{j=2}^{2^u-1}\max_{x\in I^{2^u}_j}(x^{k+1})f_c(x) $ is dominated by an integrable function $v(x)$. By applying the Lebesgue Dominated Convergence Theorem \cite{rudin1987real}, we obtain
			\begin{align*}
			\lim_{u\to\infty}\int_{\mathbb{R}}\sum_{j=2}^{2^u-1}\max_{x\in I^{2^u}_j}(x^{k})f_c(x) \ud x =\int_{\mathbb{R}}x^{k}f_c(x) \ud x.
			\end{align*}
			Since 	$$\lim_{u\to\infty}\int_{I^{2^u}_1\cup I^{2^u}_{2^u}}\max\left(\frac{Cx^{k}}{|x|^{k+1+\xi}},0\right) \ud x=0,$$
			we have
			\begin{align}
			\lim_{u\to\infty}M_{2^u}&=\lim_{u\to\infty}p_0\left(\int_{\mathbb{R}}\sum_{j=2}^{2^u-1} \max_{x\in I^{2^u}_j}(x^{k}) f(x) \ud x\right.\nonumber\\
			&\quad+\left.\int_{I^{2^u}_1\cup I^{2^u}_{2^u}}\max\left(\frac{Cx^{k}}{|x|^{k+1+\xi}},0\right) \ dx\right)+\sum_{h=1}^Hp_h\theta_h^k\nonumber\\
			&=\int_{\mathbb{R}}x^{k} f(x)\ dx +0=\E^{{f}}[X^{k}].\label{eqn:Mlimit}
			\end{align}
		\end{IEEEproof}			
	\section{Simulations}
	
\begin{figure}[!htb]
		\begin{minipage}[b]{.48\linewidth}
			\centering
			\centerline{\includegraphics[width=8.0cm]{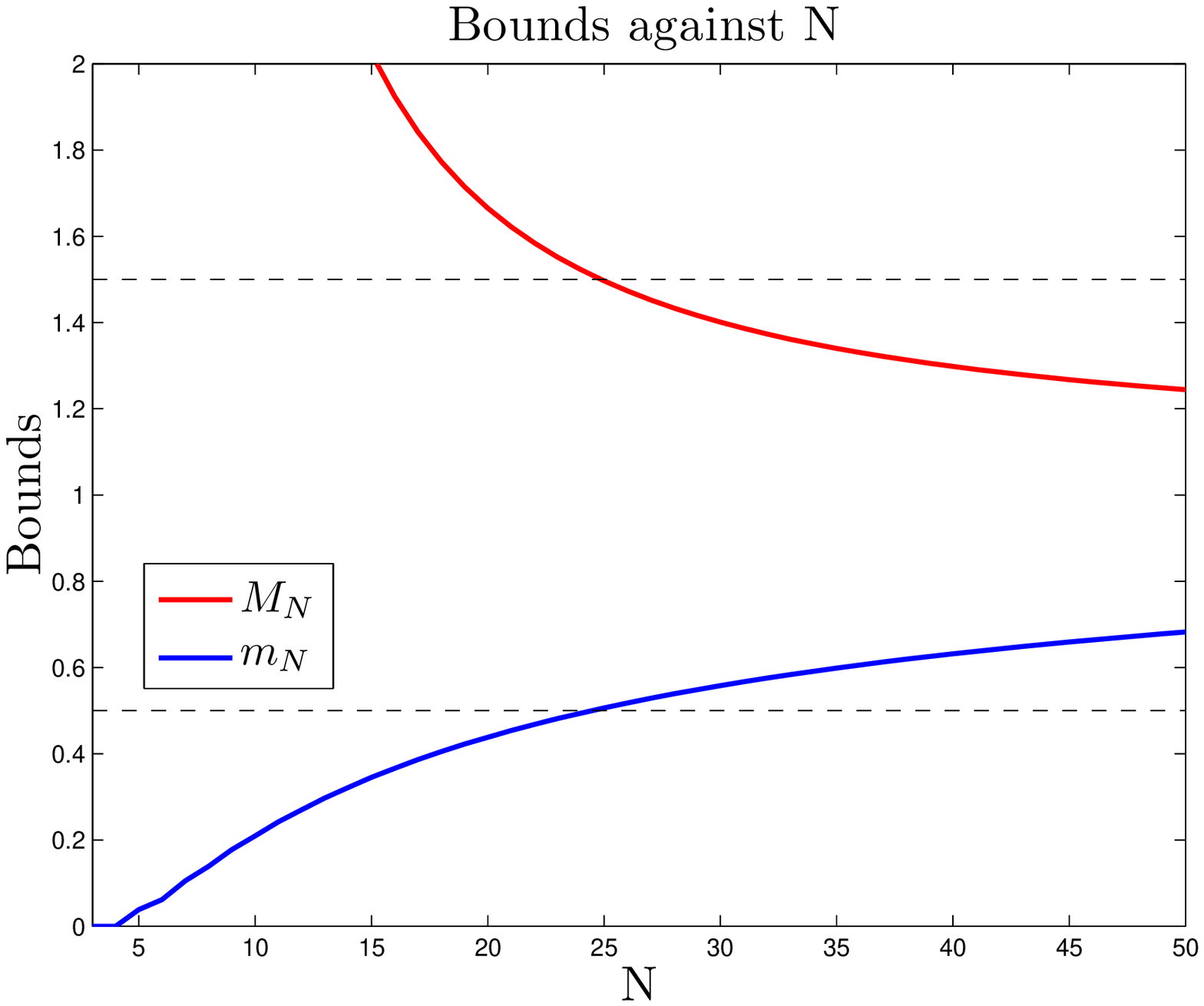}}
		  \vspace{1.5cm}
			\centerline{(a)}\medskip
		\end{minipage}
		%\hfill
		\begin{minipage}[b]{0.48\linewidth}
			\centering
		\centerline{\includegraphics[width=8.0cm]{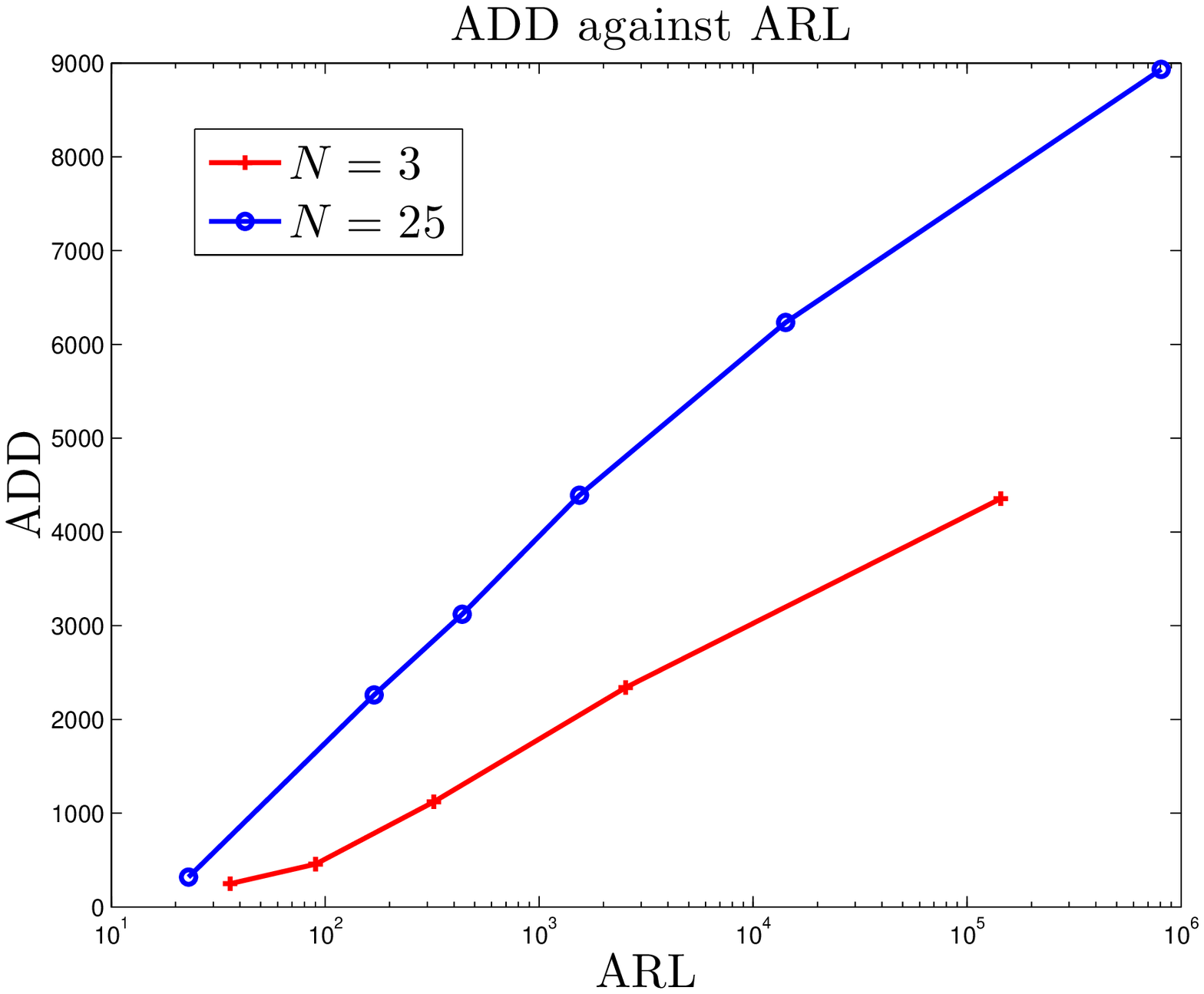}}
		  \vspace{1.5cm}
			\centerline{(b)}\medskip
		\end{minipage}
		\caption{(a) Graph of the upper and lower bounds computed in Algorithm \ref{alg:ComputeN}. (b) Comparison of ADD vs ARL performance using $N=N_{\text{exact}}$ and $N=N_{\text{approx}}$.}
		\label{fig:Exact_vs_Approx}
	\end{figure}

	We consider the case where $g=\calN(0,0.5)$, which differs in the second moment from ${f}=\calN(0,1)$.  and we compare the ADD-ARL performance of using $N=N_{\text{exact}}$ by applying Proposition \ref{prop:exact} against using $N=N_{\text{approx}}$ obtained by Algorithm \ref{alg:ComputeN}. Since the number of intersections of the cdfs of ${f}$ and $g$ is $3$, $g\in D({f},3)$. Furthermore, since both ${f}$ and $g$ are absolutely continuous symmetric distributions, $g\notin D({f},2)$. Therefore the smallest parameter $N$ such that $g\in D({f},N)$ is $N_{\text{exact}}=3$. On the other hand, applying Algorithm~\ref{alg:ComputeN} gives us an estimate for the parameter to be $N_{\text{approx}}=25$, if we assume $k=2$, $\epsilon=0.5$, $C=1.9$ and $\xi=4$. Figure \ref{fig:Exact_vs_Approx}a shows the upper bound $M_N$ and lower bound $m_N$ in Algorithm~\ref{alg:ComputeN} as the parameter $N$ varies. 
	Figure \ref{fig:Exact_vs_Approx} shows the ADD versus ARL performance for these values of $N$ using $5000$ Monte Carlo trials.
\bibliographystyle{IEEEtran}
\bibliography{IEEEabrv,StringDefinitions,refs}